\documentclass[aps,12pt,pra,superscriptaddress,showpacs,onecolumn,preprint]{article}

\usepackage{float}
\usepackage{mathtools}
\usepackage{wasysym}
\usepackage{simplewick}
\usepackage{timestamp}
\usepackage{longtable}
\usepackage{microtype}
\usepackage{lmodern}
\usepackage[T1]{fontenc}
\usepackage{color}
\usepackage{hyperref}
\usepackage{color}
\usepackage{mathtools}
\usepackage{subfigure}
\usepackage{graphicx}
\usepackage{rotating}
\usepackage{epsf,latexsym}
\usepackage{amssymb,amsmath}
\usepackage{dcolumn}
\usepackage{times}
\usepackage{amscd}
\usepackage{bm}
\usepackage{color}

\begin{document}

\title{\bf {Fluctuations in Relativistic Causal Hydrodynamics}}

\author{Avdhesh Kumar\thanks{Corresponding author. E-mail addresses: avdhesh@prl.res.in (Avdhesh Kumar), jeet@prl.res.in (Jitesh R. Bhatt), 
apmishra@prl.res.in (Ananta P. Mishra)}, Jitesh R. Bhatt, and Ananta P. Mishra}

\maketitle

\begin{center}
{\it{Theoretical Physics Division, Physical Research Laboratory, Navrangpura, Ahmedabad 380009, India}}
\end{center}

\maketitle

\begin{abstract}
 Formalism to calculate the hydrodynamic fluctuations by applying the Onsager
theory to the relativistic Navier-Stokes equation is already known. 
In this work, we calculate hydrodynamic-fluctuations within the framework of
the second order hydrodynamics of M\"{u}ller, Israel and Stewart and its generalization
to the third order. We have also calculated the fluctuations for several other
causal hydrodynamical equations.  
We show that the form for the Onsager-coefficients and form of the correlation-functions
remains same as those obtained by the relativistic Navier-Stokes equation and it 
does not depend on any specific model of hydrodynamics.
Further we numerically investigate evolution of the correlation function  
using the one dimensional boost-invariant (Bjorken) flow.
We compare the correlation functions obtained using the causal hydrodynamics with the
correlation-function for the relativistic Navier-Stokes equation.
We find that the qualitative behavior of the correlation-functions remain same 
for all the models of the causal hydrodynamics.
\end{abstract}
\noindent {\it Keywords}: Fluctuations, Onsager coefficients, Causal Hydrodynamics.
\newpage
\maketitle
\section*{1. Introduction}
A study of fluctuations in continuous media is of great interest in physics
and it can provide a link between the macroscopic and microscopic points of
view. A macroscopic theory such as hydrodynamics provides a simplest possible description
of a complicated many-body system in terms of space-time evolution of the mean or averaged
quantities like energy density, pressure, flow velocity etc. The fluctuation
theory studies small deviations from the mean behavior and it can help in calculating
correlation functions for the macroscopic variables\cite{landau,landau1}.
In context of relativistic hydrodynamics, results of the fluctuation-dissipation
theorem has been studied in Refs. \cite{salie,calzetta}. In Ref.\cite{salie} authors have 
studied the fluctuation in the contexts of general-relativistic Navier-Stokes
theory. A more general framework of hydrodynamics described as the divergence type theory(DTT)\cite{geroch}
was considered in Ref.\cite{calzetta}.
It ought to be noted that recently in an interesting work in Ref.\cite{kapusta}, 
the authors have applied results of the fluctuation-dissipation theorem to 
the relativistic {\it Navier-Stokes} theory of hydrodynamics
and calculated the two particles correlators for the one-dimensional hydrodynamics (Bjorken) flow
relevant for the relativistic heavy-ion collision experiments at RHIC and LHC. 
The authors obtained several analytical results for two particle correlation functions. 
Further, in Ref.\cite{kapusta1}, the authors have studied the effect of thermal conductivity 
on the correlation function using the Bjorken-flow.
It should be noted here that it is well-known that relativistic Navier-Stokes theory
exhibit acausal behavior and it can give rise to unphysical instabilities\cite{hiscock}.
However the causality can be restored if the terms with higher orders are included in the hydrodynamics
as indicated by the Maxwell-Cattaneo law\cite{romatschke}.
Indeed these issues do not arise in the second-order causal hydrodynamics theory developed by M\"{u}ller, Israel and
Stewart (MIS)\cite{mis}.
Form of the Navier-Stokes equations can be determined from the second law of thermodynamics $\partial_\mu S^\mu \geq 0$, 
where $S^\mu$  denotes the equilibrium entropy current\cite{landau1}. However, in general it is not possible 
for an out-of-equilibrium fluid to have an equilibrium entropy current\cite{romatschke}. In MIS 
hydrodynamics out-of-equilibrium current can have contributions from dissipative processes 
like the effect of viscosity and the heat conduction. This has an interesting analogy with the irreversible thermodynamics \cite{djou,herrera}. 
Further the MIS hydrodynamics has been extensively applied to study the relativistic heavy-ion
collisions\cite{romatschke,muronga,muronga1,heinz} and also in cosmology\cite{berera}.
Later this formalism was extended to include the effect of third order terms in the gradient expansion
\cite{ajel}. Recently, it has been shown that the derivation of the MIS equations from the underlying kinetic
equation may not be unique, there may exist a more general set of hydrodynamic equations
which may allow one to obtain MIS equations as a special case\cite{rischke,bhalerao}. 

Finally, it should be mentioned here that although the 
divergent type theory (DTT) of relativistic fluid of Geroch-Lindblom\cite{geroch} allows for a consistent proof of causality 
and stability of its solutions, it is far from direct thermodynamic intuition. 
Moreover, the connection between the DTTs and MIS or other causal hydrodynamics theories is not yet clearly established.
                      
 In this work we apply Onsager theory to MIS equations and
also to the hydrodynamics models developed by Denicol, Koide and Rischke (DKR) \cite{rischke},
Jaiswal, Bhalerao and Pal (JBP)\cite{bhalerao} and other models based on MIS approach \cite{mis,ajel,roma1}. 
Further, we apply these results to study the hydrodynamical
evolution using 1+0 dimensional Bjorken flow. In particular, we calculate the correlators using
the Onsager coefficients for various relativistic hydrodynamical theories.

\section*{2.  Fluctuations and correlations in Hydrodynamics}
In thermodynamic equilibrium entropy of the system $S$ which is a function of
the additive quantities $x_k$ becomes maximum. In equilibrium, $S$ satisfies the condition $X_k=-\frac{\partial S}{\partial x_k}=0$.
However, when the system is slightly away from the equilibrium the generalized forces $X_k\neq 0$
and $\frac{dx_i}{dt}=-\gamma_{ik}X_k+\xi_{i}$, the summation convention is implied, 
describes the flux associated with the quantity $x_i$, here $\xi_{i}$ are the random forces or 
the noise term and $\gamma_{ik}$ are the Onsager coefficients. 
The Onsager reciprocity relations imply that
$\gamma_{ik}=\gamma_{ki}$. In this phenomenological theory time rate of change of
the total entropy $\frac{dS}{dt}$ is given by, 
\begin{equation}
 \frac{d S}{dt}= -\frac{dx_i}{dt}X_i \label{1} ,
\end{equation}
which can also be written as,
\begin{equation}
 \frac{d S}{dt}= \gamma_{ik}X_{k}X_{i}-\xi_{i}X_{i} \label{2} .
\end{equation}

Correlation between $\xi_{i}$ is given by the formula, 
\begin{equation}
\langle{\xi_{i}(t_1)\xi_{k}(t_2)}\rangle=(\gamma_{ik}+\gamma_{ki})\delta(t_1-t_2).
\label{3} 
\end{equation}

 The correlation functions can be found once $\gamma_{ij}$ are known\cite{landau1,kapusta}. 
 In order to find out $\gamma_{ij}$, one needs to know the $\frac{dS}{dt}$ for the underlying   
hydrodynamical theory together with identification of the generalized forces and fluxes. The expression 
for rate of change of entropy $\frac{dS}{dt}$ can be found either by using equations
of hydrodynamics together with the thermodynamic relations or from the kinetic
theory\cite{mis}. In our work we are using the former approach. 

We start with the expressions for the energy momentum tensor
$T^{\mu\nu}$ and the current-density $J^{\mu}_B$ for a viscous fluid,
\begin{eqnarray}
T^{\mu\nu}&=&T^{\mu\nu}_{id}+{\Delta}T^{\mu\nu}+S^{\mu\nu},\\
\label{4}
J^{\mu}_{B}&=&n_{B}u^{\mu}+{\nu}^{\mu}+I^{\mu},
\label{5}
\end{eqnarray}
where, 
$T^{\mu\nu}_{id}={\epsilon}u^{\mu}u^{\nu}-p{\Delta}^{\mu\nu}$ is the ideal 
part of the energy momentum tensor with $\Delta^{\mu\nu}=g^{\mu\nu}-u^{\mu}u^{\nu}$, and 
$\epsilon$, $p$, $u^{\mu}$ are 
the local energy density, pressure and fluid flow four-velocity respectively. It is to be noted that
$u^{\mu}u_{\mu}=1$ and $g_{\mu\nu}=diag(+,-,-,-)$. ${\Delta}T^{\mu\nu}={\Delta}T^{\mu\nu}_{vis}+{\Delta}T^{\mu\nu}_{heat}$
with ${\Delta}T^{\mu\nu}_{vis}=\pi^{\mu\nu}-\Delta^{\mu\nu}\Pi$
and ${\Delta}T^{\mu\nu}_{heat}=W^{\mu}u^{\nu}+W^{\nu}u^{\mu}$, is the dissipative part 
of the energy momentum tensor and $S^{\mu\nu}$ is the stochastic term arising due
to the local thermal fluctuations\cite{kapusta}. Similarly, ${\nu}^{\mu}$ and $I^{\mu}$
denote the dissipative(non-equilibrium) and the stochastic terms in baryon current density respectively. $n_{B}$ denotes the net 
density of baryon number in local rest frame. $W^{\mu}=q^{\mu}+h{\nu}^{\mu}$ is the energy flow in 
local rest frame, $h=(\epsilon+p)/n_B$ is the enthalpy per particle 
and ${\nu}^{\mu}={\Delta}^{\mu\nu}J_B{\nu}$ is the baryon number flow in the 
local rest frame.
For the dissipative fluxes one can always require the following relations to hold
$u_{\mu}\pi^{\mu\nu}=0$,\ $\pi^{\alpha}_{\alpha}=0$,\ ${\Delta}_{\mu\nu}\pi^{\mu\nu}=0$,\
$u_{\mu}W^{\mu}=0$,\ $u_{\mu}{\nu}^{\mu}=0$,\ $u_{\mu}q^{\mu}=0$.

The relevant conservation equations for the hydrodynamics can be written as, 
\begin{eqnarray}
\label{6}
\partial_{\mu}J^{\mu}_{B}&=&{D}n_{B}+n_{B}\nabla_{\mu}u^{\mu}+\partial_{\mu}{\nu}^{\mu}=0,\\
\label{7}
u_{\nu}\partial_{\mu}T^{\mu\nu}&=&D\epsilon+(\epsilon+p+\Pi)\nabla_{\mu}u^{\mu}-
\pi_{\mu\nu}\nabla^{\langle\mu}u^{\nu\rangle}+\nabla_{\mu}W^{\mu}-2W^{\mu}Du_{\mu}=0,\\
\Delta^{\alpha}_{\nu}\partial_{\mu}T^{\mu\nu}&=&(\epsilon+p+\Pi)Du^{\alpha}-\nabla^{\alpha}(p+\Pi)
+\Delta^{\alpha\nu}\nabla^{\sigma}\pi_{\nu\sigma}-\pi^{\alpha\nu}Du_{\nu}\nonumber\\& &+\Delta^{\alpha\nu}DW_{\nu}+
2W^{(\alpha}\nabla_{\nu}u^{\nu)}=0\label{8},
\end{eqnarray}
where, $D=u^{\mu}\partial_{\mu}$ and $\nabla^{\mu}=\Delta^{\mu\nu}\partial_{\mu}$.\
There are only 5 equations written above and 14 unknowns $n_{B}$, $\epsilon$, $\Pi$, $W^{\mu}$, 
$\pi^{\mu\nu}$ and $u^{\mu}$. Therefore, 9 additional equations for dissipative fluxes are required 
to obtain the hydrodynamical solution.

There are two popular choices for  $u^\mu$: In Landau-Lifshitz frame $u^\mu$ is parallel to the energy-flow and $W^\mu=0$
which implies that $q^{\mu}=-h{\nu}^{\mu}$. Another choice is the Eckart-frame, where $u^\mu$
can be parallel to $J_B{\mu}$ and ${\nu}^{\mu}=0$ and this would
imply $W^{\mu}=q^{\mu}$. Now we shall obtain the 9 
additional equations and fluctuation correlations in 
Landau-Lifshitz and Eckart frame respectively.

\subsection*{2.1  Equations for dissipative fluxes in Landau-Lifshitz frame and fluctuation correlations in MIS}
In order to derive the $9$ additional equations one needs the expression
for the out-of-equilibrium entropy four-current. In Landau-Lifshitz frame the expression for the 
non-equilibrium entropy four-current is given 
in literature\cite{mis,muronga1,heinz} and is 
as follows,

\begin{equation}
S^{\mu}=su^{\mu}-\frac{\mu_{B}}{T}\nu^{\mu}-\left(\beta_{0}\Pi^{2}-{\beta_{1}}q_{\nu}q^{\nu}+
\beta_{2}\pi_{\nu\lambda}\pi^{\nu\lambda}\right)\frac{u^{\mu}}{2T}-\frac{{\alpha_{0}}{\Pi}q^{\mu}}{T}+
\frac{{\alpha_{1}}{\pi}^{\mu\nu}q_{\nu}}{T}. 
\label{9}
\end{equation}

$\beta_{0}$, $\beta_{1}$, $\beta_{2}$ are thermodynamic co-efficients and describe the scalar, vector and tensor contribution to the entropy 
density respectively. $\alpha_{0}$ and $\alpha_{1}$ are functions of energy density $\epsilon$ and baryon density $n_{B}$ and they describe 
viscous and heat coupling.

The divergence of the non-equilibrium 
entropy four-current (Eq.(\ref{9})) using Eqs.(\ref{6}-\ref{7}) and the thermodynamic 
relations $d\epsilon=Tds-\mu{d}n$ and $\epsilon+p=Ts-\mu{n}$ can be 
written as follows,
\begin{align}
T{\partial}_{\mu}S^{\mu}=&-\Pi\left[\partial_{\mu}u^{\mu}+\beta_{0}\dot{\Pi}+
\frac{1}{2}T{\partial}_{\mu}\left(\frac{\beta_0}{T}u^{\mu}\right)\Pi+{\alpha_{0}}\nabla_{\mu}q^{\mu}\right]\nonumber\\&-
q^{\mu}\left[-h^{-1}{T}\nabla_{\mu}\left(\frac{\mu}{T}\right)-{\beta_{1}}\dot{q}_{\mu}-\frac{1}{2}T\partial_{\nu}\left
(\frac{{\beta_{1}}}{T}u^{\nu}\right)q_{\mu}
-{\alpha_{1}}\partial_{\nu}\pi^{\nu}_{\mu}+{\alpha_{0}}\partial_{\mu}\Pi\right]\nonumber\\&+
\pi^{\mu\nu}\left[\sigma_{\mu\nu}
-\beta_{2}\dot{\pi}_{\mu\nu}-\frac{1}{2}T\partial_{\lambda}\left(\frac{\beta_{2}}{T}u^{\lambda}\right)\pi_{\mu\nu}+
{\alpha_{1}}\nabla_{\langle\nu}q_{\mu\rangle}\right],\nonumber\\ 
\label{10}
\end{align}
where we have used the notation $\dot{F}=DF$.
According to the second law of thermodynamics we must have $T\partial_{\mu}S^{\mu}\geq0$. This inequality will be 
satisfied if $\Pi$,$q_{\mu}$ and $\pi^{\mu\nu}$ satisfy the following equations, 
\begin{eqnarray}
\label{11}
\left[\partial_{\mu}u^{\mu}+\beta_{0}\dot{\Pi}+
\frac{1}{2}T{\partial}_{\mu}\left(\frac{\beta_0}{T}u^{\mu}\right)\Pi+{\alpha_{0}}\nabla_{\mu}q^{\mu}\right]&=&
-\frac{\Pi}{\zeta},\\ 
\left[-h^{-1}{T}\nabla_{\mu}\left(\frac{\mu}{T}\right)-{\beta_{1}}\dot{q}_{\mu}-\frac{1}{2}T\partial_{\nu}\left
(\frac{{\beta_{1}}}{T}u^{\nu}\right)q_{\mu}
-{\alpha_{1}}\partial_{\nu}\pi^{\nu}_{\mu}+{\alpha_{0}}\partial_{\mu}\Pi\right]&=&\frac{q_{\mu}}{\lambda{T}},\\
\label{12}
\left[\sigma_{\mu\nu}
-\beta_{2}\dot{\pi}_{\mu\nu}-\frac{1}{2}T\partial_{\lambda}\left(\frac{\beta_{2}}{T}u^{\lambda}\right)\pi_{\mu\nu}+
{\alpha_{1}}\nabla_{\langle\nu}q_{\mu\rangle}\right]&=&\frac{\pi_{\mu\nu}}{2\eta}. 
\label{13}
\end{eqnarray}

Here we note that sometimes the term with factor $1/2$ on the left hand 
side of equations (\ref{11}-\ref{13}) are ignored by arguing 
that the gradient of thermodynamic quantities are small \cite{muronga2,heinz}. But in the present 
case we are retaining these terms. Thus,
Eqs.(\ref{11}-\ref{13}) can be written as,
\begin{eqnarray}
\label{14}
 \tau_{\Pi}\dot{\Pi}+\Pi&=&-\zeta\nabla_{\alpha}u^{\alpha}-
 l_{\Pi{q}}\nabla_{\mu}q^{\mu}-\left(\frac{1}{2}T\zeta\partial_{\mu}\left(\frac{\tau_{\Pi}u^{\mu}}{{\zeta}T}\right)\right)\Pi,\\
 \tau_{q}\dot{q}_{\mu}+q_{\mu}&=& -\lambda{T^2}h^{-1}\nabla_{\mu}\left(\frac{\mu}{T}\right)+l_{q\Pi}\nabla_{\mu}\Pi-
 l_{q\pi}\nabla_{\nu}\pi^{\nu}_{\mu}+\frac{1}{2}\lambda{T^2}\partial_{\nu}\left(\frac{\tau_{\pi}u^{\nu}}{\lambda{T^2}}\right)q_{\mu},\\
 \label{15}
 \tau_{\pi}\dot{\pi}_{\mu\nu}+\pi_{\mu\nu}&=&
 2\eta\sigma_{\mu\nu}+
 l_{\pi{q}}\nabla_{\langle\mu}q_{\nu\rangle}-\eta{T}\partial_{\lambda}\left(\frac{\tau_{\pi}u^{\lambda}}{2\eta{T}}\right)\pi_{\mu\nu},
 \label{16}
 \end{eqnarray}
 Henceforth, we call Eqs.(\ref{14}-\ref{16}) as MIS equations. Here, 
 $\tau_{\Pi}=\zeta\beta_0$, $\tau_{q}={\lambda}T\beta_1$, $\tau_{\pi}=2\eta\beta_2$ 
 are identified as the relaxation times and $l_{\Pi{q}}=\zeta\alpha_0$, $l_{q\Pi}={\lambda}T\alpha_0$, $l_{q\pi}={\lambda}T\alpha_1$, 
$l_{\pi{q}}=2\eta\alpha_1$ as coupling constants. These 9 equations for the dissipative 
fluxes together with the Eqs.(\ref{6}-\ref{8}) and equation of state form complete set of the 
hydrodynamic equations. Note that the limit $\tau_{\Pi}, \tau_{q}, \tau_{\pi}, l_{\Pi{q}}, l_{q\Pi}, l_{q\pi}, l_{\pi{q}}\rightarrow0$ 
is the first order limit which correspond to the Navier-Stokes case.

Eq.(\ref{10}) can be written as,
 
\begin{equation}
T\partial_{\mu}S^{\mu}=\frac{{\Pi}^{2}}{\zeta}-\frac{q^{\mu}q_{\mu}}{{\lambda}T}+\frac{\pi^{\mu\nu}\pi_{\mu\nu}}{2\eta}\geq0.
\label{17}
\end{equation}

Here, $q^{\mu}q_{\mu}<0$\cite{muronga}.
Now using the identity $\Delta^{\mu\nu}\Delta_{\mu\nu}=3$ and the condition $\Delta_{\mu\nu}\pi^{\mu\nu}=0$,
one can write Eq.(\ref{17}) as,

\begin{equation}
\partial_{\mu}{S^{\mu}}=\frac{\Delta{T^{\mu\nu}_{vis}}}{T}\left(\frac{\pi_{\mu\nu}}{2\eta}-\frac{\Delta_{\mu\nu}\Pi}{3\zeta}\right)
-\frac{q^{\mu}q_{\mu}}{\lambda{T^2}}.\label{18}
\end{equation}
Upon integrating over the whole volume Eq.(\ref{18}) can be written as,
\begin{equation}
\frac{dS}{dt}=\int{d^{3}x}\left[\frac{\Delta{T^{\mu\nu}_{vis}}}{T}\left(\frac{\pi_{\mu\nu}}{2\eta}-\frac{\Delta_{\mu\nu}\Pi}{3\zeta}\right)
-\frac{q^{\mu}q_{\mu}}{\lambda{T^2}}\right].
\label{19}
\end{equation}
Following identification between the phenomenological variables ($\dot{x}_1$, $\dot{x}_2$) and the 
hydrodynamical variables can be made\cite{kapusta},
\begin{equation}
\dot{x_1}\rightarrow\Delta{T^{\mu\nu}_{vis}} \ , \ \dot{x_2}\rightarrow{q}^{\mu}.\label{20}
\end{equation}
A comparison of Eq.(\ref{19}) with the phenomenological equation Eq.(\ref{1}) will give,
\begin{eqnarray}
X_1&=&-\frac{1}{T}\left(\frac{\pi_{\mu\nu}}{2\eta}-\frac{\Delta_{\mu\nu}\Pi}{3\zeta}\right)\Delta{V},\\
\nonumber
X_2&=&\frac{q_{\mu}}{\lambda{T}^2}\Delta{V}.
\label{21}
\end{eqnarray}

Now neglecting the stochastic term in Eq.(\ref{2}) and comparing it with Eq.(\ref{19}) one can get,
\begin{eqnarray}
\label{22}
\gamma_{11}X_{1}&=&-\Delta{T^{\mu\nu}_{vis}},\\
\label{23}
\gamma_{22}X_{2}&=&-q^{\mu},\\
\gamma_{12}&=&\gamma_{21}=0 .
\label{24}
\end{eqnarray} 

The coefficients $\gamma_{12}$ and $\gamma_{21}$ are zero, because the dissipative fluxes due to heat 
and viscosity are considered to be independent.
Coefficients $\gamma_{11}$ and $\gamma_{22}$ are rank-four tensors and they can 
be parameterized as follows,

\begin{equation}
\gamma_{11}=\left[A\Delta^{\mu\nu\alpha\beta}+B\Delta^{\mu\nu}\Delta^{\alpha\beta}\right]\frac{1}{\Delta{V}}, \,\,\
\gamma_{22}=\frac{C\Delta^{\mu\nu}}{\Delta{V}},\label{25}
\end{equation}
where, $\Delta^{\mu\nu\alpha\beta}=\Delta^{\mu\alpha}\Delta^{\nu\beta}-\frac{1}{3}\Delta^{\mu\nu}\Delta^{\alpha\beta}$.
Now using Eqs.(\ref{22}, \ref{23}) one can determine the coefficients $A=2\eta{T}$, $B=\zeta{T}$ and $C=-\lambda{T^{2}}$. 
Thus one can write,
\begin{equation}
\gamma_{11}=2T\left[\left(\eta\Delta^{\mu\alpha}\Delta^{\nu\beta}
-\frac{1}{3}\eta\Delta^{\mu\nu}\Delta^{\alpha\beta}\right)+\frac{1}{2}\zeta\Delta^{\mu\nu}\Delta^{\alpha\beta}\right]\frac{1}{\Delta{V}},\,\,\,
\gamma_{22}=-\frac{\lambda{T^2}\Delta^{\mu\nu}}{\Delta{V}}.
\label{26}
\end{equation}

From above expression of $\gamma_{11}$ one can see that there is a additive contribution 
of shear and bulk viscosity i.e one can write it as $\gamma_{11}=(\gamma_{11})_{\eta}+(\gamma_{11})_{\zeta}$.

Now following Eq.(\ref{2}), the correlation functions can be written as,
\begin{eqnarray}
\label{27}
 \langle{S^{\mu\nu}_{vis}(x_1)}S^{\alpha\beta}_{vis}(x_2)\rangle&=&
 2T\left[\eta(\Delta^{\mu\alpha}\Delta^{\nu\beta}+\Delta^{\mu\beta}\Delta^{\nu\alpha})+
(\zeta-\frac{2}{3}\eta)\Delta^{\mu\nu}\Delta^{\alpha\beta}\right]\delta(x_1-x_2),~~~~~~~~~~~\\
 \langle{I^{\mu}(x_1)}I^{\nu}(x_2)\rangle&=&-2\lambda{T^2}\Delta^{\mu\nu}\delta(x_1-x_2),\\
 \label{28}
 \langle{S^{\mu\nu}_{vis}(x_1)}I^{\alpha}(x_2)\rangle&=&0.
  \label{29}
\end{eqnarray}
These are the stochastic or noise auto-correlation functions for the 
MIS hydrodynamics in the Landau-Lifshitz frame. 

\subsection*{2.2  Equations for dissipative fluxes in Eckart frame and fluctuation correlations in MIS}
In the Eckart frame expression for the entropy four-current\cite{mis,muronga,muronga1} can be written as,

\begin{equation}
S^{\mu}=su^{\mu}+\frac{q_{\mu}}{T}-\left(\beta_{0}\Pi^{2}-\bar{\beta}_{1}q_{\nu}q^{\nu}+
\beta_{2}\pi_{\nu\lambda}\pi^{\nu\lambda}\right)\frac{u^{\mu}}{2T}-\frac{\bar{\alpha}_{0}{\Pi}q^{\mu}}{T}+
\frac{\bar{\alpha}_{1}{\pi}^{\mu\nu}q_{\nu}}{T}. 
\label{30}
\end{equation}
Note that here coefficients $\beta_{0}, \beta_{2}$ are same as in Landau-Lifshitz case while the 
coefficients $\bar{\alpha}_{i}$ and $\bar{\beta}_{1}$ are given as,
$\bar{\alpha}_{i}=\alpha_{i}+\frac{1}{\epsilon+p}$ and $\bar{\beta}_{1}=\beta_{1}+\frac{1}{\epsilon+p}$, where 
$\alpha_{i}$, $\beta_{1}$ are the coefficients in Landau-Lifshitz frame. Next, divergence of the non-equilibrium 
entropy four-current (Eq.(\ref{30})) using Eqs.(\ref{6}-\ref{7}) and the thermodynamic 
relations $d\epsilon=Tds-\mu{d}n$ and $\epsilon+p=Ts-\mu{n}$ can be 
written as follows,
\begin{align}
T{\partial}_{\mu}S^{\mu}=&-\Pi\left[\partial_{\mu}u^{\mu}+\beta_{0}\dot{\Pi}+
\frac{1}{2}T{\partial}_{\mu}\left(\frac{\beta_0}{T}u^{\mu}\right)\Pi+\bar{\alpha}_{0}\nabla_{\mu}q^{\mu}\right]\nonumber\\&-
q^{\mu}\left[\nabla_{\mu}ln{T}-\dot{u}_{\mu}-\bar{\beta}_{1}\dot{q}_{\mu}-\frac{1}{2}T\partial_{\nu}\left
(\frac{\bar{\beta}_{1}}{T}u^{\nu}\right)q_{\mu}
-\bar{\alpha}_{1}\partial_{\nu}\pi^{\nu}_{\mu}+\bar{\alpha}_{0}\partial_{\mu}\Pi\right]\nonumber\\&+\pi^{\mu\nu}\left[\sigma_{\mu\nu}
-\beta_{2}\dot{\pi}_{\mu\nu}-\frac{1}{2}T\partial_{\lambda}\left(\frac{\beta_{2}}{T}u^{\lambda}\right)\pi_{\mu\nu}+
\bar{\alpha}_{1}\nabla_{\langle\nu}q_{\mu\rangle}\right].\nonumber\\ 
\label{31}
\end{align}
In order to have $T\partial_{\mu}S^{\mu}\geq0$ we must have the following equations for 
the dissipative fluxes,
\begin{eqnarray}
\label{32}
\left[\partial_{\mu}u^{\mu}+\beta_{0}\dot{\Pi}+
\frac{1}{2}T{\partial}_{\mu}\left(\frac{\beta_0}{T}u^{\mu}\right)\Pi+\bar{\alpha}_{0}\nabla_{\mu}q^{\mu}\right]&=&-\frac{\Pi}{\zeta},\\
\left[\nabla_{\mu}ln{T}-\dot{u}_{\mu}-\bar{\beta}_{1}\dot{q}_{\mu}-\frac{1}{2}T\partial_{\nu}\left
(\frac{\bar{\beta}_{1}}{T}u^{\nu}\right)q_{\mu}
-\bar{\alpha}_{1}\partial_{\nu}\pi^{\nu}_{\mu}+\bar{\alpha}_{0}\partial_{\mu}\Pi\right]&=&\frac{q_{\mu}}{\lambda{T}},\\
\label{33}
\left[\sigma_{\mu\nu}
-\beta_{2}\dot{\pi}_{\mu\nu}-\frac{1}{2}T\partial_{\lambda}\left(\frac{\beta_{2}}{T}u^{\lambda}\right)\pi_{\mu\nu}+
\bar{\alpha}_{1}\nabla_{\langle\nu}q_{\mu\rangle}\right]&=&\frac{\pi_{\mu\nu}}{2\eta}.
\label{34}
\end{eqnarray}

\noindent
Thus Eq.(\ref{31}) can be written as,
\begin{equation}
T\partial_{\mu}S^{\mu}=\frac{{\Pi}^{2}}{\zeta}-\frac{q^{\mu}q_{\mu}}{{\lambda}T}+\frac{\pi^{\mu\nu}\pi_{\mu\nu}}{2\eta}\geq0,
\label{35}
\end{equation}
\noindent
which can easily be casted into the following form,

\begin{equation}
T\partial_{\mu}S^{\mu}={\Delta}T^{\mu\nu}\left[\frac{\pi_{\mu\nu}}{2\eta}-\frac{\Delta_{\mu\nu}\Pi}{3\zeta}
-\frac{1}{2{\lambda}T}(u_{\nu}q_{\mu}+u_{\mu}q_{\nu})\right].
\label{36}
\end{equation}
Upon integrating over the whole volume Eq.(\ref{36}) can be written as,
\begin{equation}
\frac{dS}{dt}=\int{d}^{3}x\frac{{\Delta}T^{\mu\nu}}{T}\left[\frac{\pi_{\mu\nu}}{2\eta}-\frac{\Delta_{\mu\nu}\Pi}{3\zeta}
-\frac{1}{2{\lambda}T}(u_{\nu}q_{\mu}+u_{\mu}q_{\nu})\right],
\label{37}
\end{equation}
which can be rearranged in the following form,
\begin{equation}
\frac{dS}{dt}=\int{d}^{3}x\left[\frac{{\Delta}T^{\mu\nu}_{vis}}{T}\left(\frac{\pi_{\mu\nu}}{2\eta}-
\frac{\Delta_{\mu\nu}\Pi}{3\zeta}\right)
+\frac{{\Delta}T^{\mu\nu}_{heat}}{T}\left[\left(\frac{\pi_{\mu\nu}}{2\eta}-
\frac{\Delta_{\mu\nu}\Pi}{3\zeta}\right)-\frac{1}{2{\lambda}T}\left(u_{\nu}q_{\mu}+u_{\mu}q_{\nu}\right)\right]\right].~~~~~~~~~~
\label{38} 
\end{equation}
In this case also one can make the identifications as before,
\begin{equation}
\dot{x}_1\rightarrow\Delta{T^{\mu\nu}_{vis}} \ , \ \dot{x}_2\rightarrow\Delta{T^{\mu\nu}_{heat}}.\label{39}
\end{equation}
The comparison between Eqs. (\ref{38}) and (\ref{1}) will give,
\begin{eqnarray}
X_1&=&-\frac{1}{T}\left(\frac{\pi_{\mu\nu}}{2\eta}-\frac{\Delta_{\mu\nu}\Pi}{3\zeta}\right)\Delta{V},\\
\nonumber
 X_2&=&-\frac{1}{T}\left[\left(\frac{\pi_{\mu\nu}}{2\eta}-\frac{\Delta_{\mu\nu}\Pi}{3\zeta}\right)
 -\frac{1}{2\lambda{T}}(u_{\nu}q_{\mu}+u_{\mu}q_{\nu})\right]\Delta{V}.
 \label{40}
\end{eqnarray}
Again neglecting the stochastic term in Eq.(\ref{2}) and comparing it with Eq.(\ref{38}) one can get,
\begin{eqnarray}
\label{41}
\gamma_{11}X_{1}&=&-\Delta{T^{\mu\nu}_{vis}},\\
\label{42}
\gamma_{22}X_{2}&=&-\Delta{T^{\mu\nu}_{heat}},\\
 \gamma_{12}&=&\gamma_{21}=0.
 \label{43}
\end{eqnarray}
One can use the following parameterization for $\gamma_{11}$ and $\gamma_{22}$,
\begin{equation}
\gamma_{11}=\left[A\Delta^{\mu\nu\alpha\beta}+B\Delta^{\mu\nu}\Delta^{\alpha\beta}\right]\frac{1}{\Delta{V}}, \,\,\
\gamma_{22}=\Big[\bar{A}\Delta^{\mu\alpha}u^{\nu}u^{\beta}+\bar{B}\Delta^{\nu\beta}u^{\mu}u^{\alpha}\Big]\frac{1}{\Delta{V}}.\label{44}
\end{equation}

Since we know the forms of $(X_{1},X_{2})$ and $(\Delta{T^{\mu\nu}_{vis}},\Delta{T^{\mu\nu}_{heat}})$, therefore using  Eqs.(\ref{44}) 
and Eqs.(\ref{41}-\ref{42}), one can determine the coefficients 
$A=2\eta{T}$, $B=\zeta{T}$ and $\bar{A}=\bar{B}=-2\lambda{T^2}$.
Thus $\gamma_{11}$ and $\gamma_{22}$ can be written as,
\begin{eqnarray}
\gamma_{11}&=&2T\left[\left(\eta\Delta^{\mu\alpha}\Delta^{\nu\beta}
-\frac{1}{3}\eta\Delta^{\mu\nu}\Delta^{\alpha\beta}\right)+\frac{1}{2}\zeta\Delta^{\mu\nu}\Delta^{\alpha\beta}\right]\frac{1}{\Delta{V}},\\
\label{45}
\gamma_{22}&=&-2\lambda{T^2}\left[\Delta^{\mu\alpha}u^{\nu}u^{\beta}+
\Delta^{\nu\beta}u^{\mu}u^{\alpha}\right]\frac{1}{\Delta{V}}.
\label{46}
\end{eqnarray}
Thus one can write the correlation functions using Eq.(\ref{3}) as,
\begin{eqnarray}
\label{47}
  \langle{S^{\mu\nu}_{vis}(x_1)}S^{\alpha\beta}_{vis}(x_2)\rangle&=&
 2T\left[\eta(\Delta^{\mu\alpha}\Delta^{\nu\beta}+\Delta^{\mu\beta}\Delta^{\nu\alpha})+
(\zeta-\frac{2}{3}\eta)\Delta^{\mu\nu}\Delta^{\alpha\beta}\right]\delta(x_1-x_2),~~~~~~~\\
 \langle{S^{\mu\nu}_{heat}(x_1)}S^{\alpha\beta}_{heat}(x_2)\rangle&=&
 -2\lambda{T^2}\Big[\Delta^{\mu\alpha}u^{\nu}u^{\beta}+
\Delta^{\nu\beta}u^{\mu}u^{\alpha}+\Delta^{\mu\beta}u^{\nu}u^{\alpha}\nonumber\\&+&
\Delta^{\nu\alpha}u^{\mu}u^{\beta}\Big]\delta(x_1-x_2),~~~~~~~~\\
\label{48}
\langle{{S^{\mu\nu}_{vis}(x_1)}S^{\alpha\beta}_{heat}(x_2)}\rangle&=&0.
 \label{49}
\end{eqnarray}

Form of these correlations is very similar to the correlations 
obtained for the relativistic Navier-Stokes case\cite{kapusta}. The relaxation time for the dissipative fluxes do not appear 
explicitly in the expressions for the correlation. However, the evolution 
of the correlations can be very different as demonstrated later.

\subsection*{2.3  Equations for dissipative fluxes in Landau-Lifshitz frame and fluctuation correlations for other Hydrodynamic models}
\vspace{0.5cm}
In this section we consider some of the interesting alternate approaches to the causal  MIS hydrodynamics
and some of its extensions.

\subsubsection*{2.3.1  Third order hydrodynamics}
In the Ref.\cite{ajel} third order corrections to the MIS hydrodynamics was considered when 
the effect of bulk-viscosity and heat-flux were absent. In this case expression for the non-
equilibrium entropy four-current can be written as,

\begin{equation}
S^{\mu}=su^{\mu}-\frac{\beta_{2}}{2{T}}{\pi_{\alpha\beta}}{\Pi^{\alpha\beta}u^{\mu}}+
\alpha\frac{\beta^{2}_{2}}{T}\pi_{\alpha\beta}\pi^{\alpha}_{\sigma}\pi^{\beta\sigma}u^{\mu},\label{50}
\end{equation}
where, $\alpha$ is a new dimensionless coefficient and it is assumed to be a constant. The last 
term on the right hand side of the above equation represents the third order correction to 
the equation of entropy. In order to fulfill the requirement of maximal entropy at equilibrium, the third 
order term must satisfy the condition $\alpha\frac{\beta^{2}_{2}}{T}\pi_{\alpha\beta}\pi^{\alpha}_{\sigma}\pi^{\beta\sigma}u^{\mu}\leq0$.
Divergence of the entropy four-current can be written as,
\begin{align}
 \partial_{\mu}S^{\mu}=&\frac{1}{T}\pi_{\alpha\beta}\sigma^{\alpha\beta}
 -\pi_{\alpha\beta}\pi^{\alpha\beta}\partial_{\mu}\left(\frac{\beta_{2}}{2T}u^{\mu}\right)
 -\frac{\beta_{2}}{T}\pi_{\alpha\beta}\dot{\pi}^{\alpha\beta}\nonumber\\&+
 \alpha\partial_{\mu}\left(\frac{\beta^{2}_{2}}{T}u^{\mu}\right)\pi_{\alpha\beta}\pi^{\alpha}_{\sigma}{\pi}^{\beta\sigma}
 +3\tau_{\pi}\theta\alpha\frac{\beta^{2}_{2}}{T}\pi_{\alpha\beta}\pi^{\alpha}_{\sigma}\dot{\pi}^{\beta\sigma}\geq0.\label{51}
\end{align}
Here, the Knudsen number(=$\tau_{\pi}\theta$) is required to satisfy the condition $\tau_{\pi}\theta\ll1$ for the 
validity of hydrodynamic approach. For the condition $T\partial_{\mu}S^{\mu}\geq0$ to be satisfied one must have, 
\begin{equation}
\partial_{\mu}s^{\mu}=
\frac{1}{2\eta{T}}\pi^{\mu\nu}\pi_{\mu\nu},\label{52}
\end{equation}
which implies that the form of shear viscous tensor $\pi^{\alpha\beta}$ should be given by,
\begin{align}
 \pi^{\alpha\beta}=2\eta{T}\left[\frac{1}{T}\sigma^{\alpha\beta}
 -\pi^{\alpha\beta}\partial_{\mu}\left(\frac{\beta_{2}}{2T}u^{\mu}\right)
 -\frac{\beta_{2}}{T}\dot{\pi}^{\alpha\beta}+
 \alpha\partial_{\mu}\left(\frac{\beta^{2}_{2}}{T}u^{\mu}\right)\pi^{\alpha}_{\sigma}{\pi}^{\beta\sigma}
 +3\tau_{\pi}\theta\alpha\frac{\beta^{2}_{2}}{T}\pi^{\alpha}_{\sigma}\dot{\pi}^{\beta\sigma}\right].\label{53}
\end{align}

Since $\tau_{\pi}\theta\sim\frac{\tau_{\pi}}{\tau}$ is of same order as $\frac{\pi^{\alpha\beta}}{T^{4}}$, therefore,  
the last term in above equation is a fourth order term\cite{ajel}. Thus neglecting the last term one can write above equation as,

\begin{align}
\dot{\pi}^{\alpha\beta}=-\frac{\pi^{\alpha\beta}}{\tau_{\pi}}+\frac{\sigma^{\alpha\beta}}{\beta_{2}}
 -\pi^{\alpha\beta}\frac{T}{\beta_{2}}\partial_{\mu}\left(\frac{\beta_{2}}{2T}u^{\mu}\right)
 +\alpha\frac{T}{\beta_{2}}\partial_{\mu}\left(\frac{\beta^{2}_{2}}{T}u^{\mu}\right)\pi^{\alpha}_{\sigma}{\pi}^{\beta\sigma}
 .\label{54}
\end{align}

Here coefficient $\alpha$ and $\beta_{2}$ have respectively the values $\frac{8}{3}$ and $\frac{9}{4e}$ as given in Ref.\cite{ajel}.

Now starting from Eq.(\ref{52}) and following similar prescription to determine the Onsager coefficient as 
done in second-order MIS hydrodynamics one can write
\begin{equation}
 \dot{x}=\pi^{\mu\nu} \ , \ X=-\frac{1}{2\eta{T}}{\pi^{\mu\nu}}\Delta{V},\label{55} 
\end{equation}
and the Onsager coefficient,
\begin{equation}
\gamma=2\eta{T}\left[\Delta^{\mu\alpha}\Delta^{\nu\beta}-\frac{1}{3}\Delta^{\mu\nu}\Delta^{\alpha\beta}\right]\frac{1}{\Delta{V}}.
\label{56}
\end{equation}
The viscous correlation function can be written as,
\begin{equation}
\langle{S^{\mu\nu}_{vis}(x_1)}S^{\alpha\beta}_{vis}(x_2)\rangle=
 2T\left[\eta(\Delta^{\mu\alpha}\Delta^{\nu\beta}+\Delta^{\mu\beta}\Delta^{\nu\alpha})
-\frac{2}{3}\eta\Delta^{\mu\nu}\Delta^{\alpha\beta}\right]\delta(x_1-x_2).
\label{57}
\end{equation}
One can notice that this expression is same as the one obtained using the 
 second-order theory with $\Pi=0$.

\subsubsection*{2.3.2  JBP hydrodynamics}

In the Ref\cite{bhalerao} the authors have constructed the expression for the entropy 
four-current $S^{\mu}$ generalized from the Boltzmann's H-function and find out the expression 
for its divergence as,
\begin{align}
\partial_{\mu}S^{\mu}=&-\frac{\Pi}{T}\left[\theta+\beta_{0}\dot{\Pi}+\beta_{\Pi\Pi}\Pi\theta+\alpha_{0}\nabla_{\mu}n^{\mu}+
\psi\alpha_{n\Pi}n_{\mu}\dot{u}^{\mu}+\psi\alpha_{\Pi{n}}n_{\mu}\nabla^{\mu}\alpha\right]\nonumber\\&
-\frac{{n}^{\mu}}{T}\Big[T\nabla_{\mu}\alpha-\beta_{1}\dot{n}^{\mu}-\beta_{nn}n_{\mu}\theta+\alpha_{0}\nabla_{\mu}\Pi
+\alpha_{1}\nabla_{\nu}\pi^{\nu}_{\mu}+\tilde{\psi}\alpha_{n\Pi}\Pi\dot{u}_{\mu}\nonumber\\&+\tilde{\psi}\alpha_{\Pi{n}}\Pi\nabla_{\mu}\alpha
+\tilde{\chi}\alpha_{{\pi}n}\pi^{\nu}_{\mu}\nabla_{\nu}\alpha+\tilde{\chi}\alpha_{n{\pi}}\pi^{\nu}_{\mu}\dot{u}_{\nu}\Big]\nonumber\\&
+\frac{\pi^{\mu\nu}}{T}\left[\sigma_{\mu\nu}-\beta_{2}\dot{\pi}_{\mu\nu}-\beta_{\pi\pi}\theta\pi_{\mu\nu}-\alpha_{1}\nabla_{\langle\mu}n_{\nu\rangle}
-\chi\alpha_{{\pi}n}n_{\langle\mu}\nabla_{\nu\rangle}\alpha-\chi\alpha_{n\pi}n_{\langle\mu}\dot{u}_{\nu\rangle}\right]\nonumber\\
\label{58}
\end{align}
where $\theta=\partial_{\mu}u^{\mu}$. 
The second law of thermodynamics $T\partial_{\mu}S^{\mu}\geq0$ is guaranteed to be satisfied 
if we have,
\begin{equation}
T\partial_{\mu}S^{\mu}=\frac{{\Pi}^{2}}{\zeta}-\frac{n^{\mu}n_{\mu}}{\lambda}+\frac{\pi^{\mu\nu}\pi_{\mu\nu}}{2\eta},
\label{59}
\end{equation}
therefore, $\pi$, $n^{\mu}$ and $\pi^{\mu\nu}$ should satisfy the following equations,
\begin{eqnarray}
\Big[\theta+\beta_{0}\dot{\Pi}+\beta_{\Pi\Pi}\Pi\theta+\alpha_{0}\nabla_{\mu}n^{\mu}+
\psi\alpha_{n\Pi}n_{\mu}\dot{u}^{\mu}+\psi\alpha_{\Pi{n}}n_{\mu}\nabla^{\mu}\alpha\Big]&=&-\frac{\Pi}{\zeta},\\
\label{60}
\Big[T\nabla_{\mu}\alpha-\beta_{1}\dot{n}^{\mu}-\beta_{nn}n_{\mu}\theta+\alpha_{0}\nabla_{\mu}\Pi
+\alpha_{1}\nabla_{\nu}\pi^{\nu}_{\mu}+\tilde{\psi}\alpha_{n\Pi}\Pi\dot{u}_{\mu}\nonumber\\
+\tilde{\psi}\alpha_{\Pi{n}}\Pi\nabla_{\mu}\alpha
+\tilde{\chi}\alpha_{{\pi}n}\pi^{\nu}_{\mu}\nabla_{\nu}\alpha+\tilde{\chi}\alpha_{n{\pi}}\pi^{\nu}_{\mu}\dot{u}_{\nu}\Big]&=&\frac{n^{\mu}}{\lambda},\\
\label{61}
\Big[\sigma_{\mu\nu}-\beta_{2}\dot{\pi}_{\mu\nu}-
 \beta_{\pi\pi}\theta\pi_{\mu\nu}-\alpha_{1}\nabla_{\langle\mu}n_{\nu\rangle}
-\chi\alpha_{{\pi}n}n_{\langle\mu}\nabla_{\nu\rangle}\alpha-\chi\alpha_{n\pi}n_{\langle\mu}\dot{u}_{\nu\rangle}\Big]&=&\frac{\pi^{\mu\nu}}{2\eta},
\label{62}
\end{eqnarray}
where $\lambda,\zeta,\eta\geq0$ are the coefficient of charge conductivity, 
bulk viscosity and shear viscosity respectively. Coefficients $\alpha_{i}, \beta_{i}, \alpha_{XY}, \beta_{XX}$ 
are the additional transport coefficients and the parameters 
$\psi,\chi$ along with $\tilde{\psi}=1-\psi$ and $\tilde{\chi}=1-\chi$ describe the contributions 
due to the cross terms of $\Pi$ and $\pi^{\mu\nu}$ with ${n}^{\mu}$.\\

Onsager coefficients in this case too, can be obtained using the parameterization[see Eq.(\ref{25})],  
\begin{equation}
\gamma_{11}=2T\left[\eta\Delta^{\mu\alpha}\Delta^{\nu\beta}+
\frac{1}{2}(\zeta-\frac{2}{3}\eta)\Delta^{\mu\nu}\Delta^{\alpha\beta}\right]\frac{1}{\Delta{V}},
\label{63}
\end{equation}
\begin{equation}
\gamma_{22}=-\frac{\lambda{T}\Delta^{\mu\nu}}{\Delta{V}}.
\label{64}
\end{equation}
It should be noted that in Ref.\cite{bhalerao} the authors have used $\frac{n^{\mu}}{\lambda}$ 
in Eq.(\ref{59}) instead of $\frac{n^{\mu}}{\lambda{T}}$ and therefore Onsager coefficient differ 
by factor $T$ (see for example, Eqs.(\ref{26} and \ref{64})).
The correlation functions can be written as,
\begin{eqnarray}
\label{65}
 \langle{S^{\mu\nu}_{vis}(x_1)}S^{\alpha\beta}_{vis}(x_2)\rangle&=&
 2T\left[\eta(\Delta^{\mu\alpha}\Delta^{\nu\beta}+\Delta^{\mu\beta}\Delta^{\nu\alpha})+
(\zeta-\frac{2}{3}\eta)\Delta^{\mu\nu}\Delta^{\alpha\beta}\right]\delta(x_1-x_2),~~~~~~~~~~~~~~\\
 \langle{I^{\mu}(x_1)}I^{\nu}(x_2)\rangle&=&-2\lambda{T}\Delta^{\mu\nu}\delta(x_1-x_2),~~~~~~~~~~~\\
\label{66}
 \langle{S^{\mu\nu}_{vis}(x_1)}I^{\alpha}(x_2)\rangle&=&0.
 \label{67}
\end{eqnarray}

\subsubsection*{2.3.3  DKR hydrodynamics}

 In Ref.\cite{rischke}, it was demonstrated that derivation of relativistic viscous 
hydrodynamic equation from the 14-moment method done by Israel and Stewart 
may not be unique. In Ref.\cite{rischke}, authors obtained relativistic dissipative hydrodynamic 
equations for the dissipative fluxes as,
\begin{equation}
\dot{\Pi}=-\frac{\Pi}{\tau_{\Pi}}-\beta_{\Pi}\theta
-\delta_{\Pi\Pi}\Pi\theta+\lambda_{\Pi\pi}\pi^{\mu\nu}\sigma_{\mu\nu},
\label{68}
\end{equation}
\begin{equation}
\dot{\pi}^{\langle\mu\nu\rangle}=-\frac{\pi^{\mu\nu}}{\tau_{\pi}}+2\beta_{\pi}\sigma^{\mu\nu}+
2\pi_{\alpha}^{\langle\mu}\omega^{\nu\rangle\alpha}-\delta_{\pi\pi}\pi^{\mu\nu}\theta-
\tau_{\pi\pi}\pi_{\alpha}^{\langle\mu}\sigma^{\nu\rangle\alpha}+\lambda_{\pi\Pi}\Pi\sigma^{\mu\nu},
\label{69}
\end{equation}
where $\theta=\nabla_{\alpha}u^{\alpha}$, and ${\tau}'s, {\beta}'s, {\delta}'s, {\lambda}'s$ are the 
transport coefficients.

It should be noted that Eq.(\ref{69}) contains  vorticity 
term $\omega^{\alpha\beta}=\frac{1}{2}(\nabla^{\alpha}u^{\beta}-\nabla^{\beta}u^{\alpha})$. 
Note that in writing the above equations we have considered the fluid with no net baryon number. Thus the 
Eq.(\ref{7}) with no net baryon number can be written as,
\begin{equation}
\partial_{\mu}(su^{\mu})=\frac{\pi^{\mu\nu}\sigma_{\mu\nu}}{T}-\frac{\Pi\nabla_{\alpha}u^{\alpha}}{T},\label{70}
\end{equation}
From Eq.(\ref{68}) and (\ref{69}) it is easy to write,
\begin{equation}
{\nabla_{\alpha}u^{\alpha}}=-\frac{\dot{\Pi}}{\beta_{\Pi}}-\frac{\Pi}{\beta_{\Pi}\tau_{\Pi}}-
\frac{\delta_{\Pi\Pi}\Pi\nabla_{\alpha}u^{\alpha}}{\beta_{\Pi}}+\frac{\lambda_{\Pi\pi}\pi^{\mu\nu}\sigma_{\mu\nu}}{\beta_{\Pi}},\label{71}
\end{equation}
\begin{equation}
\sigma_{\mu\nu}=\frac{{\dot{\pi}^{\langle\mu\nu\rangle}}}{2\beta_{\pi}}+\frac{\pi^{\mu\nu}}{2\beta_{\pi}\tau_{\pi}}
-\frac{\pi^{\langle\mu}_{\alpha}\omega^{\nu\rangle\alpha}}{\beta_{\pi}}
+\frac{\delta_{\pi\pi}\pi^{\mu\nu}\nabla_{\alpha}u^{\alpha}}{2\beta_{\pi}}+
\frac{\tau_{\pi\pi}\pi^{\langle\mu}_{\alpha}\sigma^{\nu\rangle\alpha}}{2\beta_{\pi}}-
\frac{\lambda_{\pi\Pi}\Pi\sigma^{\mu\nu}}{2\beta_{\pi}}.\label{72}
\end{equation}

Now substituting Eq.(\ref{71}) and (\ref{72}) in Eq.(\ref{70}) one can write,

\begin{align}
\partial_{\mu}(su^{\mu})=&\frac{\pi_{\mu\nu}}{T}\left[\frac{{\dot{\pi}^{\langle\mu\nu\rangle}}}{2\beta_{\pi}}+
\frac{\pi^{\mu\nu}}{2\beta_{\pi}\tau_{\pi}}
-\frac{\pi^{\langle\mu}_{\alpha}\omega^{\nu\rangle\alpha}}{\beta_{\pi}}
+\frac{\delta_{\pi\pi}\pi^{\mu\nu}\nabla_{\alpha}u^{\alpha}}{2\beta_{\pi}}+
\frac{\tau_{\pi\pi}\pi^{\langle\mu}_{\alpha}\sigma^{\nu\rangle\alpha}}{2\beta_{\pi}}-
\frac{\lambda_{\pi\Pi}\Pi\sigma^{\mu\nu}}{2\beta_{\pi}}\right]\nonumber\\&
-\frac{\Pi}{T}\left[-\frac{\dot{\Pi}}{\beta_{\Pi}}-\frac{\Pi}{\beta_{\Pi}\tau_{\Pi}}-
\frac{\delta_{\Pi\Pi}\Pi\nabla_{\alpha}u^{\alpha}}{\beta_{\Pi}}+\frac{\lambda_{\Pi\pi}\pi^{\mu\nu}\sigma_{\mu\nu}}{\beta_{\Pi}}\right],\label{73}
\end{align}

After substituting back for $\nabla_{\alpha}u^{\alpha}$ and $\sigma_{\mu\nu}$ again from Eq.(\ref{71}) and (\ref{72}) 
into Eq.(\ref{73}) one can see the terms with the coefficients ${\delta}'s, {\tau}'s,$ and ${\lambda}'s$ are of $O(\pi^{3})$ 
or of the higher order, therefore, one can neglect these terms.

One can easily show that $\dot{\pi}^{\langle\mu\nu\rangle}=\dot{\pi}^{\mu\nu}+\pi^{\mu}_{\beta}u^{\nu}Du^{\beta}
+\pi^{\nu}_{\alpha}u^{\mu}Du^{\alpha}$. This would imply that,
\begin{equation}
\pi_{\mu\nu}\dot{\pi}^{\langle\mu\nu\rangle}=\pi_{\mu\nu}\dot{\pi}^{\mu\nu}.\label{74} 
\end{equation}

Now neglecting the the terms with the coefficients ${\delta}'s, {\tau}'s,$ and ${\lambda}'s$ from Eq.(\ref{73}) for the 
reason mentioned above, using Eq.(\ref{74}) and the identity,  
${\pi_{\mu\nu}}{{\pi^{\langle\mu}_{\alpha}\omega^{\nu\rangle\alpha}}}=0$, 
one can get,

\begin{align}
\partial_{\mu}S^{\mu}=&
\left[-\partial_{\mu}\left(\frac{u^{\mu}}{4\beta_{\pi}T}\right)+\frac{1}{2\beta_{\pi}\tau_{\pi}T}
\right]\pi^{\alpha\beta}\pi_{\alpha\beta}+
\left[-\partial_{\mu}\left(\frac{u^{\mu}}{2\beta_{\Pi}T}\right)+\frac{1}{\beta_{\Pi}\tau_{\Pi}T}
\right]\Pi^{2},
\label{75}
\end{align}
where, $S^{\mu}$ is the non-equilibrium entropy current for DKR hydrodynamics and has the form,
\begin{align}
S^{\mu}=\Bigg(su^{\mu}-\frac{\pi^{\alpha\beta}\pi_{\alpha\beta}u^{\mu}}{4\beta_{\pi}T}-
\frac{\Pi^{2}u^{\mu}}{2\beta_{\Pi}T}.........\Bigg).
\label{76}
\end{align}

Note that $\beta_{{\pi},{\Pi}}=\frac{\eta}{\tau_{{\pi},{\Pi}}}$. In Eq.(\ref{75}) the terms  
with gradients of velocity field can be neglected as
$\partial_{\mu}\left(\frac{u^{\mu}}{4{\beta_{\pi}T}}\right)=\frac{\tau_{{\pi},{\Pi}}\theta}{\eta{T}}<<\frac{1}{\eta{T}}$, where 
$\theta=\partial_{\mu}u^{\mu}$ is the inverse of the expansion scale and $\tau_{{\pi},{\Pi}}$ is relaxation time scale. For the 
the system to be in the relaxation regime, one must have $\tau_{{\pi},{\Pi}}\theta<<1$
(see Ref\cite{heinz,ajel}). Therefore from Eq.(\ref{75}) one obtains,  

\begin{equation}
\frac{dS}{dt}=\int{d^{3}x}\left[\left(\frac{1}{2\beta_{\pi}\tau_{\pi}T}\right)\pi^{\alpha\beta}\pi_{\alpha\beta}+
\left(\frac{1}{\beta_{\Pi}\tau_{\Pi}T}\right)\Pi^{2}\right]. \label{77}
\end{equation}
Further Eq.(\ref{77}) can be written in the following form,
\begin{equation}
\frac{dS}{dt}=\int{d^{3}x}\left[\frac{\Delta T^{\alpha\beta}_{vis}}{T}\left(\frac{\pi_{\alpha\beta}}{2\beta_{\pi}\tau_{\pi}}-
\frac{\Delta_{\alpha\beta}\Pi}{3\beta_{\Pi}\tau_{\Pi}}\right)\right]. \label{78}
\end{equation}
A comparison of the above expression with the phenomenological equation (Eq.(\ref{1})) yields,
\begin{equation}
\dot{x}\rightarrow{\Delta T^{\alpha\beta}_{vis}} \ , \ X\rightarrow-\frac{1}{T}\left[
\left(\frac{\pi_{\alpha\beta}}{2\beta_{\pi}\tau_{\pi}}-
\frac{\Delta_{\alpha\beta}\Pi}{3\beta_{\Pi}\tau_{\Pi}}\right)\right]\Delta{V}.\label{79}
\end{equation}

Again by comparing Eq.(\ref{78}) with Eq.(\ref{2})(when $\xi=0$) one can get,
\begin{equation}
\gamma{X}=-{\Delta T^{\mu\nu}_{vis}}.\label{80} 
\end{equation}
Where $\gamma$ is a rank four tensor and can be written as,
\begin{equation}
 \gamma=2{T}\left[\beta_{\pi}\tau_{\pi}\Delta^{\mu\nu\alpha\beta}+
 \frac{1}{2}\beta_{\Pi}\tau_{\Pi}\Delta^{\mu\nu}\Delta^{\alpha\beta}\right]\frac{1}{\Delta{V}}, \label{81}
\end{equation}
Thus the viscous correlations are,
\begin{equation}
 \langle{S^{\mu\nu}_{vis}(x_1)}S^{\alpha\beta}_{vis}(x_2)\rangle=
 2{T}\left[\beta_{\pi}\tau_{\pi}(\Delta^{\mu\alpha}\Delta^{\nu\beta}+\Delta^{\mu\beta}\Delta^{\nu\alpha})+
 (\beta_{\Pi}\tau_{\Pi}-\frac{2}{3}\beta_{\pi}\tau_{\pi})\Delta^{\mu\nu}\Delta^{\alpha\beta}\right]\delta(x_1-x_2).
\label{82}
\end{equation}

\subsubsection*{2.3.4  Conformal viscous hydrodynamics}
The entropy current for the conformal hydrodynamics\cite{roma1} can be written as, 
\begin{equation}
 S^{\mu}=\left(su^{\mu}-\frac{\tau_{\Pi}}{4\eta{T}}{\Pi_{\alpha\beta}}{\Pi^{\alpha\beta}u^{\mu}}\right).
 \label{83}
\end{equation}
 One can easily find the following expression for the Onsager coefficient and the correlation 
 function,
\begin{equation}
\gamma=2\eta{T}\left[\Delta^{\mu\alpha}\Delta^{\nu\beta}-
\frac{1}{3}\Delta^{\mu\nu}\Delta^{\alpha\beta}\right]\frac{1}{\Delta{V}},
\label{84}
\end{equation}
\begin{equation}
 \langle{S^{\mu\nu}_{vis}(x_1)}S^{\alpha\beta}_{vis}(x_2)\rangle=
 2\eta{T}\left[(\Delta^{\mu\alpha}\Delta^{\nu\beta}+\Delta^{\mu\beta}\Delta^{\nu\alpha})
-\frac{2}{3}\Delta^{\mu\nu}\Delta^{\alpha\beta}\right]\delta(x_1-x_2).
\label{85}
\end{equation}

\section*{3.  Calculation of correlation functions in boost-invariant Hydrodynamics}
As an example we apply the results obtained in the previous sections to the relativistic heavy-ion 
collisions for the Bjorken flow. According to Bjorken scenario in heavy ion collisions,  
the reaction volume is strongly expanded in the longitudinal direction, i.e along the collision 
axis(z-axis). So one can assume that there is no transverse flow. Thus one can describe flow in $1+0$ 
dimension\cite{bjorken}. It is useful to introduce the light cone variable $y$ 
and proper time $\tau$ which are defined by,
\begin{equation}
\tau=\sqrt{t^2-z^2}\,\,\,  and \,\,\,y=arc\tanh (z/t)=\frac{1}{2}\ln(\frac{t+z}{t-z}).
 \label{86}
\end{equation}
The partial derivatives in time and space can be expressed as,
\begin{equation}
\begin{bmatrix} \partial_t  \\ \partial_z \\ \end{bmatrix}=
\begin{bmatrix} \cosh y&-\sinh y \\ -\sinh y&\cosh y\end{bmatrix}=
\begin{bmatrix} \partial_{\tau}  \\ \frac{1}{\tau}\partial_y \\ \end{bmatrix}.
\label{87}
\end{equation} 
The flow velocity under the scaling assumption can be written as, 
$u^{\mu}=\gamma(1,0,0,v_{z})=(\frac{t}{\tau},0,0,\frac{z}{\tau})=(\cosh y,0,0,\sinh y)$.
We consider only longitudinal flow fluctuations and parameterize the flow velocity\cite{kouno} as,  
\begin{equation}
u^{\mu}=(\cosh\bar{\theta},\sinh\bar{\theta}), 
\label{88}
\end{equation}
where $\bar\theta=y+\delta\bar{\theta}(y,\tau)$ and $\delta\bar{\theta}(y,\tau)$ are the fluctuations 
in the longitudinal flow. In scaling limit, $\bar\theta=y$.
With this parameterization and using the transformation of derivatives one can introduce the 
operators $D$, $\nabla$ such that,
\begin{equation}
\begin{bmatrix} D  \\ \nabla \\ \end{bmatrix}=\begin{bmatrix} \cosh (\bar{\theta}-y)&\sinh (\bar{\theta}-y) \\ 
\sinh (\bar{\theta}-y)&\cosh (\bar{\theta}-y)
\end{bmatrix}=\begin{bmatrix} \partial_{\tau}  \\ \frac{1}{\tau}\partial_y \\ \end{bmatrix}.
\label{89}
\end{equation}
In the scaling limit, $D=u^{\mu}\partial_{\mu}=\frac{\partial}{\partial\tau}=\partial_\tau$ and 
$\partial_{\mu}u^{\mu}=\nabla\bar{\theta}=\frac{1}{\tau}$.\\ 
Since $S^{\mu\nu}$ satisfies the condition,
\begin{equation}
 u_{\mu}S^{\mu\nu}=0.\label{90}
\end{equation}
One can write $S^{\mu\nu}$ as\cite{kapusta},
\begin{equation}
S^{\mu\nu}=w(\tau)f(y,\tau)\Delta^{\mu\nu},
\label{91}
\end{equation}
where, $w=\epsilon+p=Ts$ and $f$ is a dimensionless quantity which satisfy
$\langle{f}\rangle=0$, where $\langle\rangle$ denotes the `average value'. In heavy-ion collision
experiments at LHC or RHIC a baryon free quark-gluon plasma is expected to be produced, 
therefore $q^{\mu}=0$. Thus only viscous-correlations are of interest, which for MIS, JBP and
third order(TO) can be written as,
\begin{equation}
 \langle{f(y_{1},\tau_{1})}{f(y_{2},\tau_{2})}\rangle=\frac{2{T(\tau_{1})}}{A\tau_{1}w^{2}(\tau_{1})}
 \left[\frac{4}{3}\eta(\tau_{1})+\zeta(\tau_{1})\right]\delta(\tau_{1}-\tau_{2})\delta(y_{1}-y_{2}),
 \label{92}
\end{equation}
where $\delta(x_{1}-x_{2})_{Transverse}$ is replaced by effective transverse 
area $A$ of colliding nuclei. Note that these correlations are same as that obtained by authors in Ref.\cite{kapusta} for 
Navier-Stokes case.
Similarly, for DKR case one can write the viscous correlations as,
\begin{equation}
\langle{f(y_{1},\tau_{1})}{f(y_{2},\tau_{2})}\rangle=\frac{2{T(\tau_{1})\left(\frac{4}{3}\beta_{\pi}\tau_{\pi}+
\beta_{\Pi}\tau_{\Pi}\right)}}{A\tau_{1}w^{2}(\tau_{1})}
\delta(\tau_{1}-\tau_{2})\delta(y_{1}-y_{2}).
 \label{93} 
\end{equation}
By defining $\eta=\beta_{\pi}\tau_{\pi}$ as in Ref.\cite{rischke} and neglecting the bulk viscosity for the correlation 
functions, one can rewrite the correlations for all the models of hydrodynamics that considered here as,

\begin{equation}
 \langle{f(y_{1},\tau_{1})}{f(y_{2},\tau_{2})}\rangle=\frac{X(\tau_{1})_{[E]}}{A}\delta(\tau_{1}-\tau_{2})\delta(y_{1}-y_{2}),
 \label{94}
\end{equation}
where,
\begin{equation}
 X(\tau_{1})_{[E]}=\frac{8}{3\tau_{1}w(\tau_{1})}
 \left(\frac{\eta(\tau_{1})}{s(\tau_{1})}\right)_{[E]}.
 \label{95}
\end{equation}
Here, subscript $[E]$ denotes the particular type of hydrodynamics model considered 
from the set of hydrodynamics models, for example $[E]=[MIS,JBP,DKR,TO,NS]$.

It is useful to study the correlation function normalized by the initial value of
the correlation obtained using the Navier-Stokes theory i.e. $C(\tau)_{[E]}=\frac{w^2(\tau) X(\tau)_{[E]}}{w^2(\tau_0)_{NS} X(\tau_0)_{NS}}$
where, $\tau_0$ is the initial-time for the hydrodynamics. $C(\tau)_{[E]}$ can also be written as, 
\begin{equation}
 C(\tau)_{[E]}=
\,\frac{\left(\frac{\tau_{0}}{\tau}\right)\left(\frac{\eta(\tau)}{s(\tau)}\right)_{[E]}\frac{w(\tau)}{w(\tau_{0})}}
 {\left(\frac{\eta(\tau)}{s(\tau)}\right)_{NS}}.
 \label{96}
\end{equation}
Further, we neglect the effect of bulk-viscosity by considering the initial temperature $T_i$
to be much larger than the critical temperature, $T_c=0.190$ GeV. 
Now, in the Landau-Lifshitz frame, the energy and the momentum conservation laws are given by,
\begin{equation}
u_{\nu}\partial_{\mu}T^{\mu\nu}=D\epsilon+(\epsilon+p)\nabla\bar{\theta}-
\pi_{\mu\nu}\nabla^{\langle\mu}u^{\nu\rangle}-S_{\mu\nu}\nabla^{(\mu}u^{\nu)}=0, 
\label{97}
\end{equation}
\begin{equation}
\Delta^{\alpha}_{\nu}\partial_{\mu}T^{\mu\nu}=(\epsilon+p)Du^{\alpha}-\nabla^{\alpha}p
+\Delta^{\alpha\nu}\nabla^{\sigma}\pi_{\nu\sigma}-\pi^{\alpha\nu}Du_{\nu}+\Delta^{\alpha\nu}\partial^{\sigma}S_{\sigma\nu}=0, 
\label{98}
\end{equation}
where, $\pi^{\alpha\beta}$ is the shear stress tensor and the dynamical equation for $\pi^{\alpha\beta}$ can 
be different for different models of hydrodynamics.

In the scaling limit $\bar{\theta}=y$, $D=u^{\mu}\partial_{\mu}=\partial_{\tau}$, 
$\partial_{\mu}u^{\mu}=\nabla\bar{\theta}=\frac{1}{\tau}$.
Using these, one can write the above equations as\cite{rischke},  
\begin{equation}
\partial_{\tau}\epsilon=-\frac{(\epsilon+p)}{\tau}+\frac{\pi}{\tau}.\label{99}
\end{equation}

Here, $\pi=\pi^{00}-\pi^{zz}$, and the noise term is considered to be smaller than the background quantities.

Now equation for $\pi$ in the scaling limit, for DKR and JBP hydrodynamics can be written  
as,
\begin{equation}
\partial_{\tau}\pi+\frac{\pi}{\tau_{\pi}}=\beta_{\pi}\frac{4}{3\tau}-\lambda\frac{\pi}{\tau}.\label{100}
\end{equation}

For JBP case, coefficients $\beta_{\pi}$, $\tau_{\pi}$ and $\lambda$ are as follows,

\begin{equation}
\beta_{\pi}=\frac{2p}{3},\ \tau^{-1}_{\pi}=\frac{5}{9}\frac{\sigma{p}}{T}, \ \lambda=4/3, \label{101}
\end{equation}
\noindent 
where, $\sigma$ is the total cross-section\cite{rischke} and it is assumed to be independent of energy\cite{huovinen,ajel}. 
For DKR hydrodynamics, the parameters $\beta_{\pi}$, $\tau_{\pi}$\cite{rischke} and $\lambda$\cite{jaiswal} are,
\begin{equation}
\beta_{\pi}=\frac{4p}{5},\ \tau^{-1}_{\pi}=\frac{3}{5}\frac{\sigma{p}}{T}, \ \lambda\equiv\frac{1}{3}\tau_{\pi\pi}+\delta_{\pi\pi}=\frac{38}{21}. 
\label{102} 
\end{equation}
Similarly equations for $\pi$ in the scaling limit MIS and third order hydrodynamics respectively can be written as,
\begin{equation}
\partial_{\tau}\pi+\frac{\pi}{\tau_{\pi}}=\frac{\eta}{\tau_{\pi}}\frac{4}{3\tau}-
\frac{1}{2}\pi\left(\frac{1}{\tau}+
{\frac{\eta{T}}{\tau_{\pi}}}\frac{\partial}{\partial{\tau}}\left(\frac{\tau_{\pi}}{\eta{T}}\right)\right).\label{103}
\end{equation}
\begin{equation}
\partial_{\tau}\pi+\frac{\pi}{\tau_{\pi}}={\frac{\eta}{\tau_{\pi}}}\frac{4}{3\tau}-
\frac{4}{3}\frac{\pi}{\tau}-\frac{\pi^{2}}{p\tau}.\label{104}
\end{equation}
Where $\frac{\eta}{\tau_{\pi}}=\frac{2p}{3}$ and $\tau^{-1}_{\pi}=\frac{5}{9}\frac{\sigma{p}}{T}$.

In what follows we consider the ideal equation of state, $\epsilon=3p$ with the pressure is given by the
bag model, $p=\frac{\pi^{2}}{30}T^{4}$. Further, we consider the initial temperature $T_i=0.310$ GeV 
and initial viscous stress $\pi$ either zero or
has the Navier-Stoke value value that is $\pi=\frac{4}{3}\frac{\eta}{\tau}$ for all the causal 
hydrodynamics and numerically solve Eqs. (\ref{99},\ref{100}), Eqs. (\ref{99},\ref{103}) and Eqs. (\ref{99},\ref{104})
for evaluating 
the correlations (\ref{96}) in case of MIS, JBP, DKR and Third order (TO) hydrodynamics. However, for the Navier-Stokes hydrodynamics 
one needs to solve only Eq.(\ref{99}) with same value of initial temperature and the viscous stress is given by,    
\begin{equation}
\pi=\eta\frac{4}{3\tau}. \label{105}
\end{equation}

The results of the numerical work are presented in the following section.

\section*{4.  Results and Discussions}

We have studied fluctuations in various models of  relativistic causal viscous hydrodynamics. 
Eqs.(\ref{27}-\ref{29}),(\ref{47}-\ref{49}), (\ref{57}), (\ref{65}-\ref{67}), (\ref{82}) and (\ref{85}) 
represent our main results describing the correlation functions for 
various models of relativistic causal hydrodynamics.
First we should like to note here that the form of the correlation functions given by 
Eqs.(\ref{27}-\ref{29}),(\ref{47}-\ref{49}), (\ref{57}), (\ref{65}-\ref{67}), (\ref{82}) and (\ref{85})  
are strikingly similar to the correlation functions obtained using relativistic Navier-Stokes theory
\cite{kapusta,salie}. The correlations do not explicitly depend upon the relaxation
times that appear in the causal theories of hydrodynamics. 
This indicates a kind of universality of the correlations given by equations  
(\ref{27}-\ref{29}),(\ref{47}-\ref{49}), (\ref{57}), (\ref{65}-\ref{67}), (\ref{82}) and (\ref{85}).
One can notice from Eqs.(\ref{27}) that the viscous correlation depends on $\epsilon+p-
\mu n$ and the ratio of viscous coefficients to the entropy density. 
The universality can be understood by the positivity argument of four entropy current i.e. 
$T\partial_{\mu}S^{\mu}=\frac{{\Pi}^{2}}{\zeta}-\frac{q^{\mu}q_{\mu}}{{\lambda}T}+
\frac{\pi^{\mu\nu}\pi_{\mu\nu}}{2\eta}\geq0$. Which is used to write the expression 
for $\frac{ds}{dt}$ by using the following properties of dissipative flues: 
$\Delta_{\mu\nu}\pi^{\mu\nu}=0$, $q_\mu u^\mu=0$ and $u_\mu\pi^{\mu\nu}=0$. These 
constraints are universal and satisfied in case of Navier-Stokes as well as all causal 
hydrodynamics no matter what form of $\pi_{\mu\nu}$,  $q^{\mu}$ and $\Pi$ is. The 
determination of Onsager coefficients [using Eq.(\ref{2})] also depends on these 
constraints leading to same form for all hydrodynamic theories and consequently 
the correlation function remains same for all theories. But in case of DTT kind 
of hydrodynamics, it is not clear if divergence of the
entropy four-current can be expressed directly in terms of scalar product of the viscosity and
heat-flux tensors. 

In order to understand the evolution of the correlation functions in some details
we have calculated the normalized correlation functions given by Eqn.(\ref{96})
for an expanding one-dimensional boost-invariant (Bjorken) flow. 
In this case all
the correlations are proportional to $\left(\epsilon+p\right)/\tau$. 
However, the details of temporal evolution of $\epsilon+p$ varies with the choice of different 
hydrodynamical models.  
In figures (1-2), we plot the normalized correlation function $C(\tau)_{[E]}$ (Eqn.\ref{96}) 
as a function of time $\tau$, where 
[E] stands for MIS, JBP, DKR, TO (Third order) and NS
hydrodynamics. Each figure has five kind of curves:
the solid (red) color curve describes the Navier-Stokes case while the dotted-dashed (blue),
the dashed (purple), the dotted (green) and large-dashed (black) curves respectively describe 
MIS, JBP, DKR and TO cases.The left panel shows the case when the initial value for
the viscous stress $\pi=0$, while the right panel represents the case when the initial
value of $\pi$ same as the Navier-Stoke case.

\begin{figure}[H]
\centering
\subfigure[]{\includegraphics[width= 8.0cm]{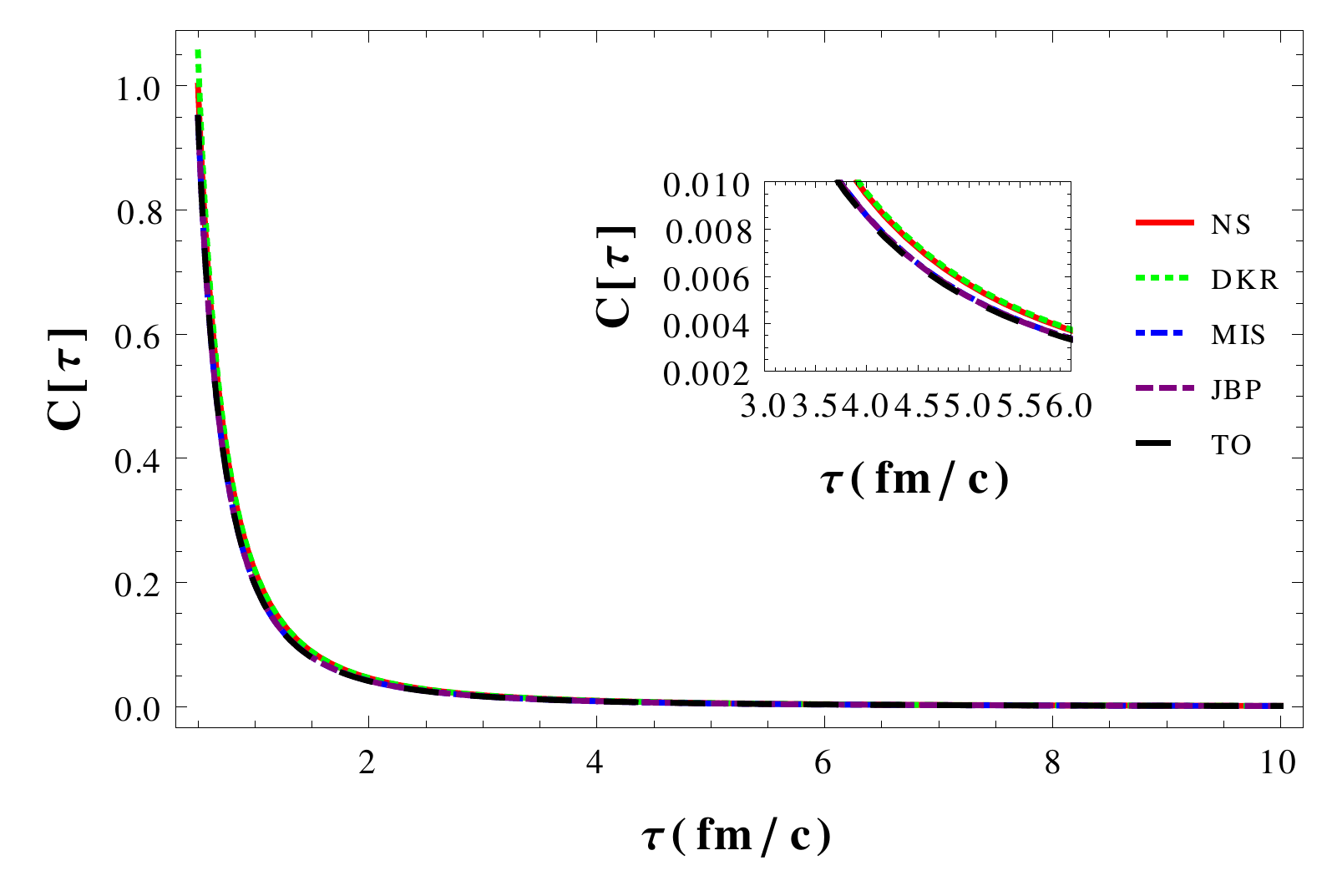}}\hspace{0.0cm}
\subfigure[]{\includegraphics[width= 8.0cm]{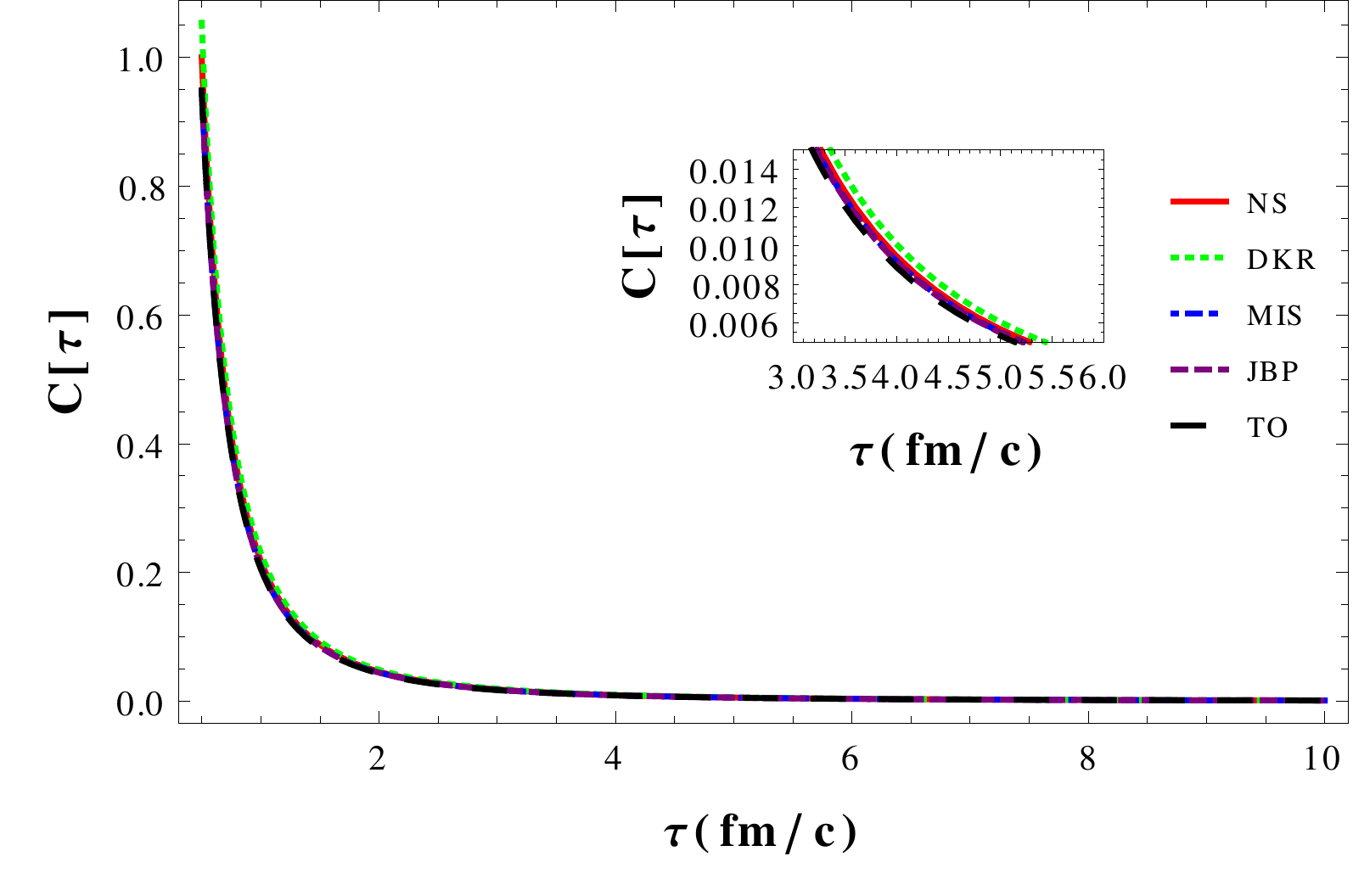}}\vspace{0.0cm}
\subfigure[]{\includegraphics[width= 8.0cm]{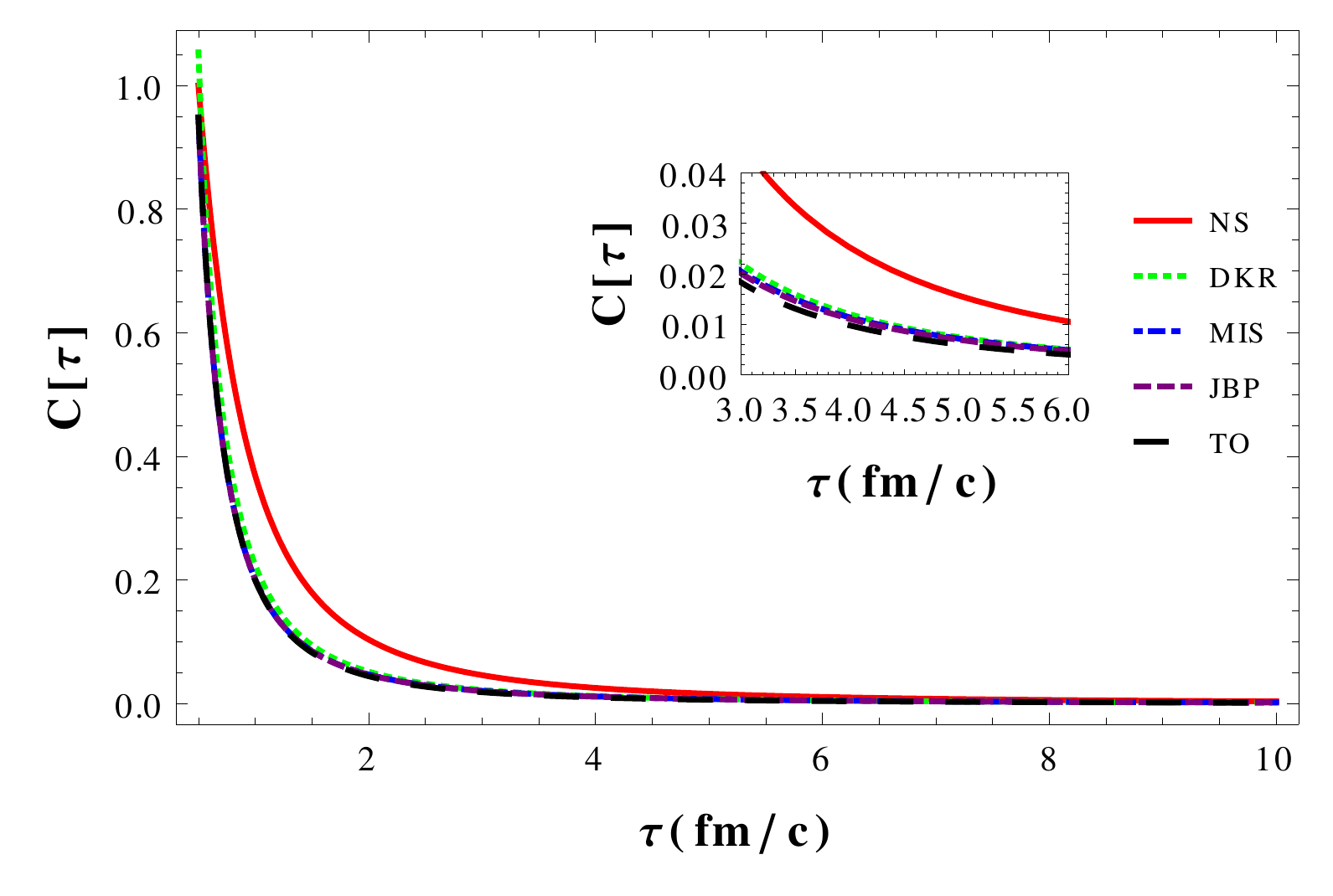}}\vspace{0.0cm}
\subfigure[]{\includegraphics[width= 8.0cm]{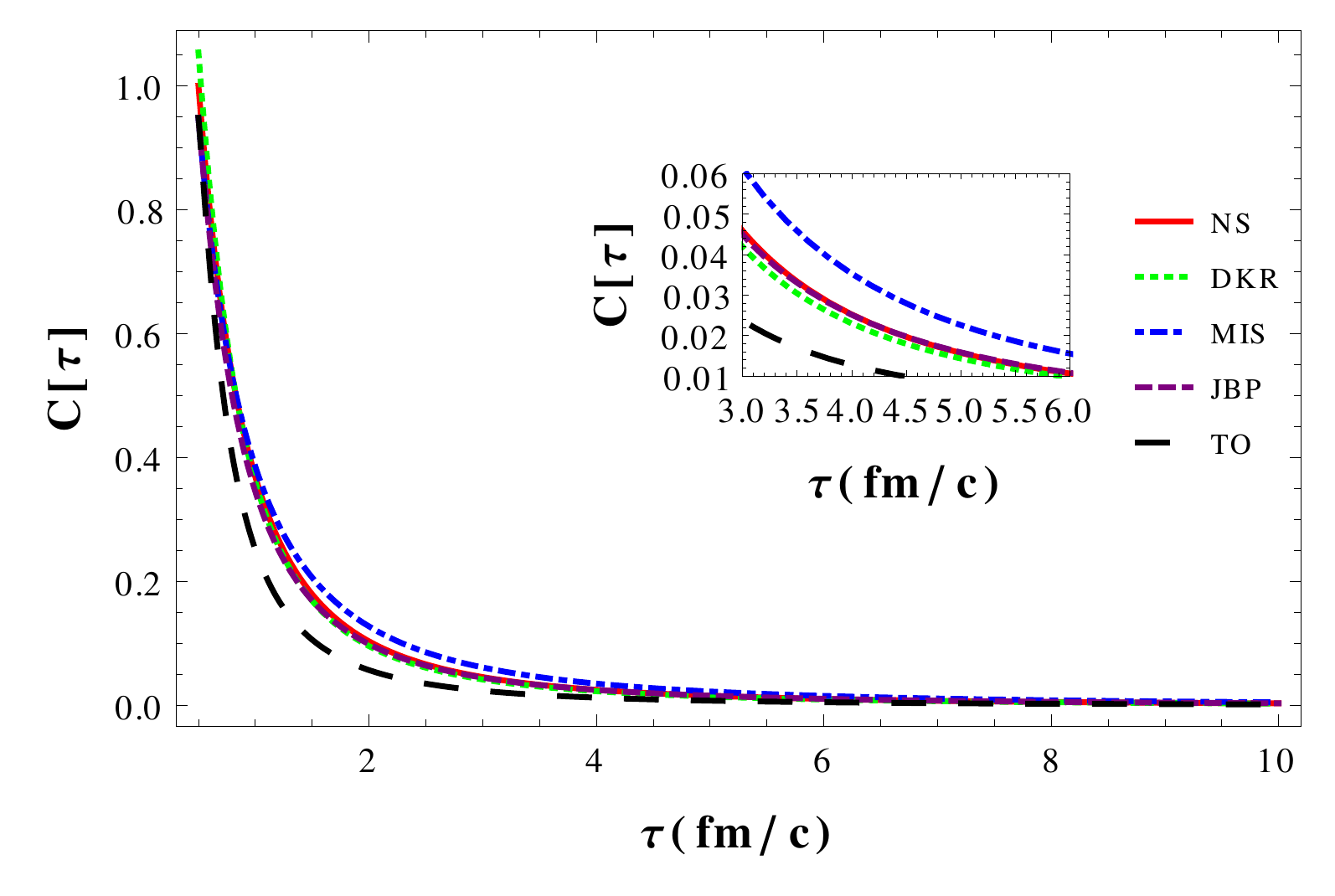}}\hspace{0.0cm}
\subfigure[]{\includegraphics[width= 8.0cm]{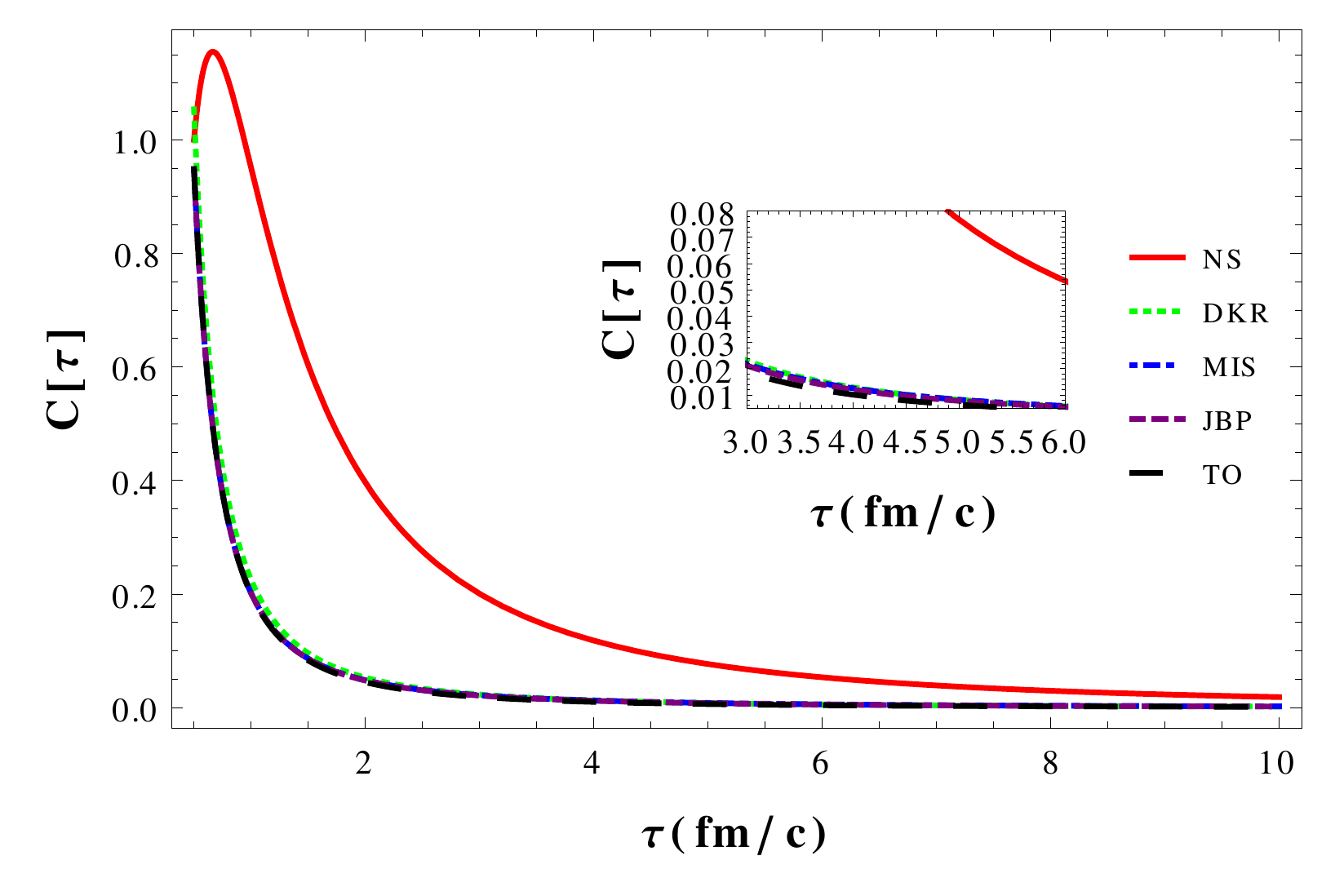}}\hspace{0.0cm}
\subfigure[]{\includegraphics[width= 8.0cm]{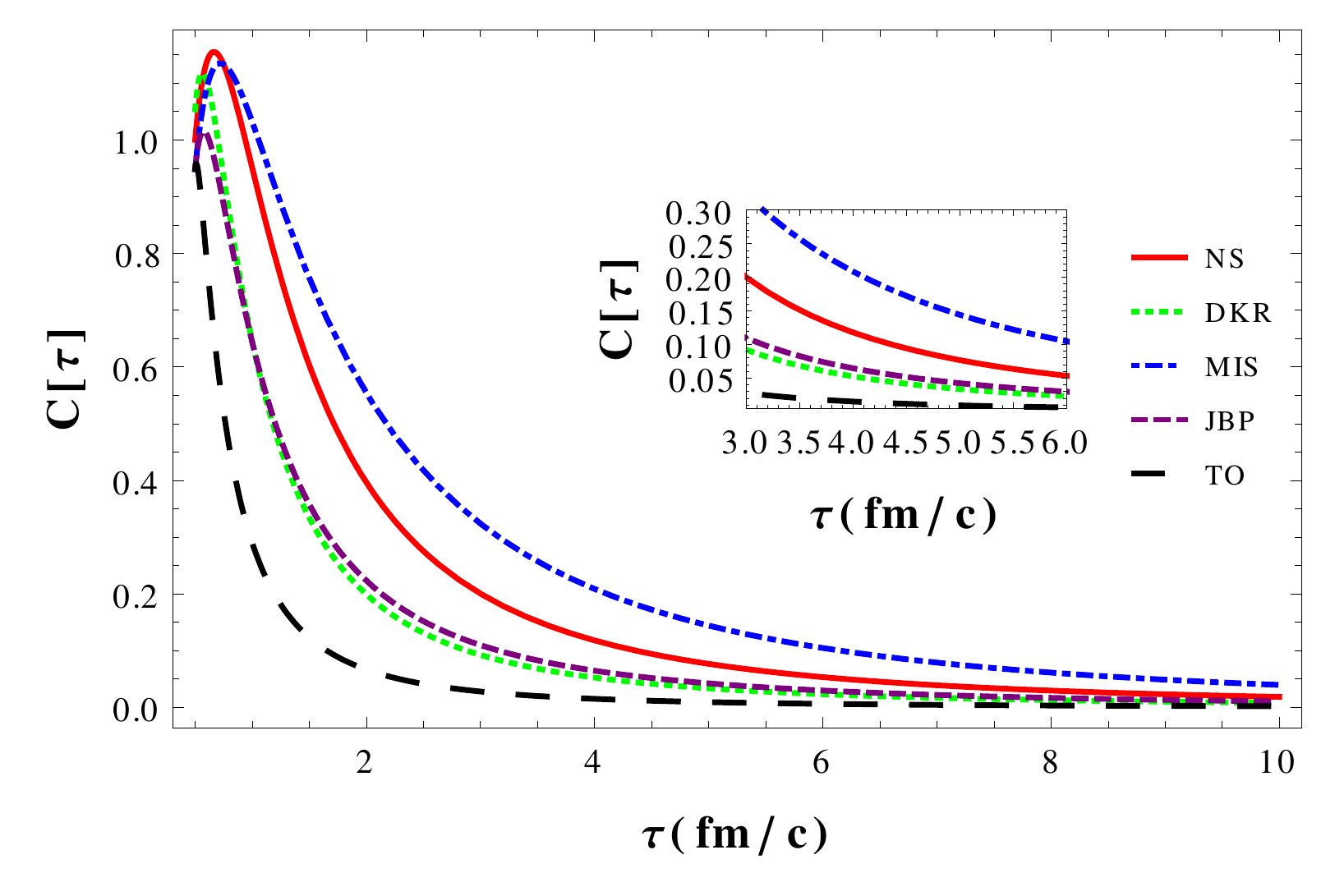}}\vspace{0.0cm}
\caption{(a), (b), (c), (d), (e) and (f) show time evolution of the function $C(\tau)_{[E]}$ [see Eq.(\ref{96})] 
with same initial temperature $T_i=310~MeV$. Where, [E] corresponds to 
NS, MIS, DKR, JBP and TO hydrodynamics. The coefficient of viscosity is calculated using 
$\eta_{DKR}=\frac{4T}{3\sigma}$. The scaling $\eta_{MIS}=\eta_{JBP}=\eta_{TO}=9/10\eta_{DKR}$ and 
$\eta_{NS}=\frac{7.59}{8}\eta_{DKR}$ ensure that the cross-section remains same in the comparison 
between the models of hydrodynamics. Cases (a), (c) and (e) corresponds to $\frac{\eta_{DKR}}{s}=0.08,\,\ 0.56,\,\ 1.60$ 
respectively with initial time $\tau_0=0.5 fm/c$  and $\pi_0=0.0$ for all causal approaches. While the 
cases (b), (d), and (f) corresponds to same $\frac{\eta_{DKR}}{s}$ and $\tau_0$ as in the former cases
But with $\pi_0$ equal to Navier-Stokes initial value for all the hydrodynamic approaches.} 
\label{fig[1]}
\end{figure} 

\begin{figure}[H]
\centering
\subfigure[]{\includegraphics[width= 8.0cm]{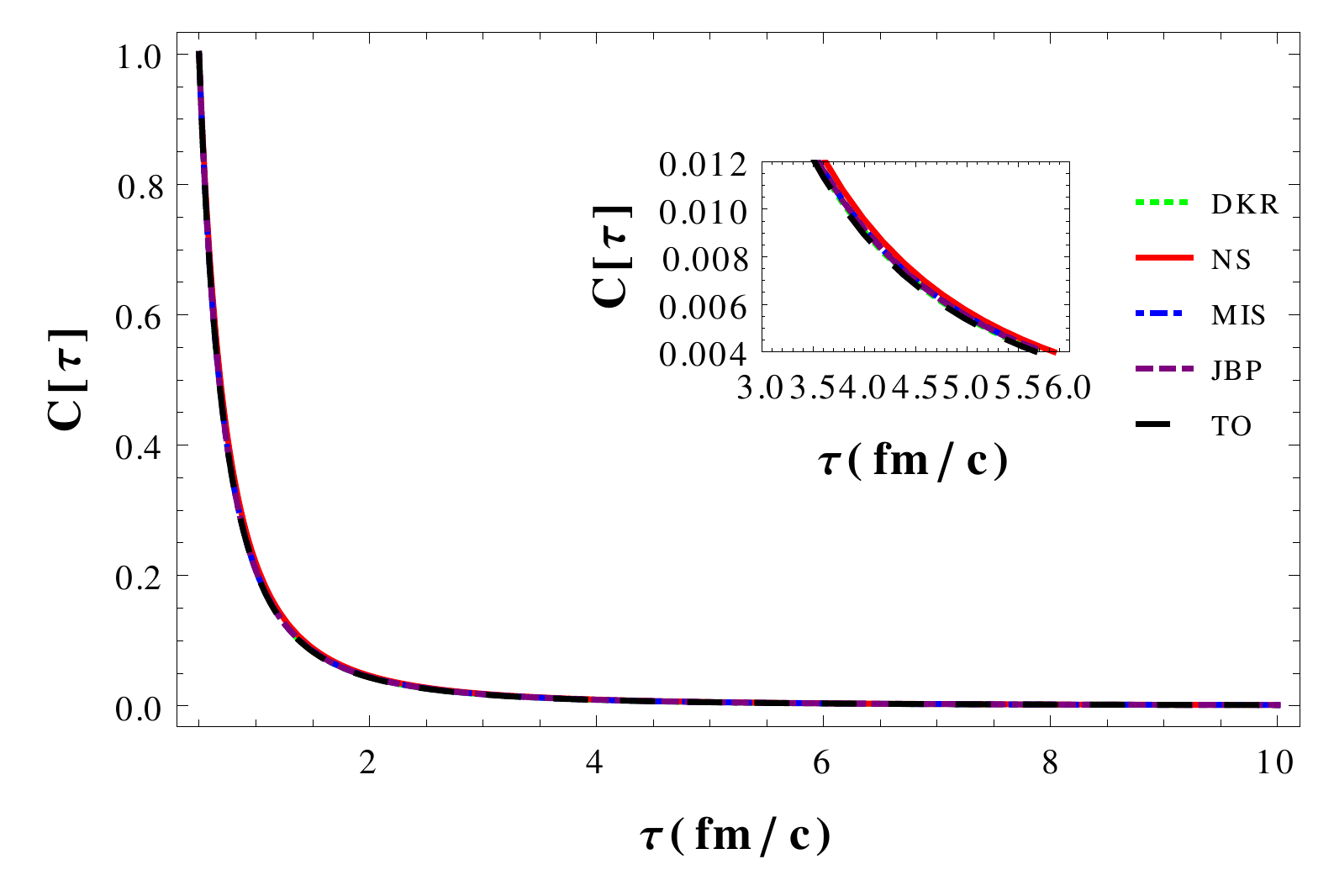}}\hspace{0.0cm}
\subfigure[]{\includegraphics[width= 8.0cm]{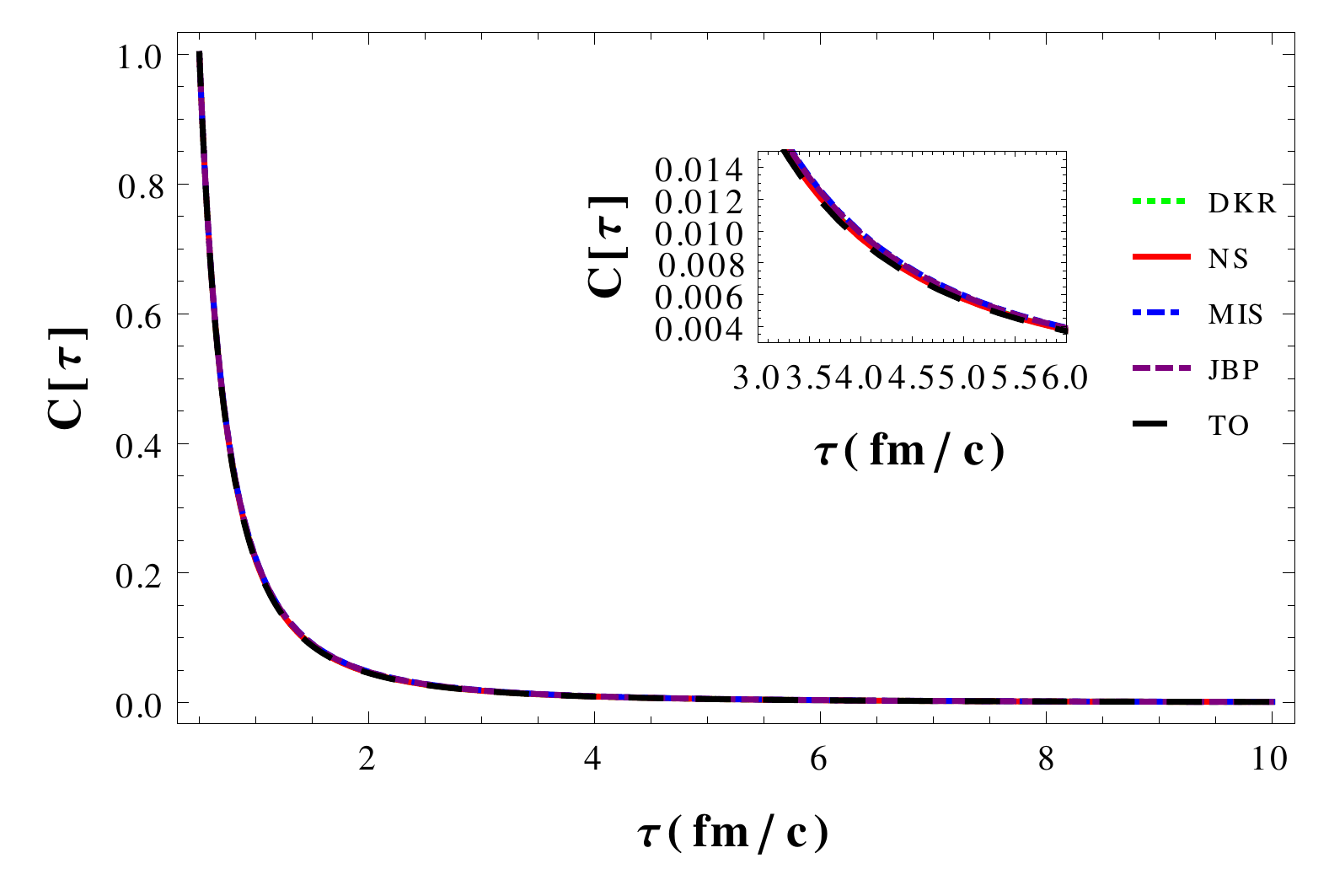}}\vspace{0.0cm}
\subfigure[]{\includegraphics[width= 8.0cm]{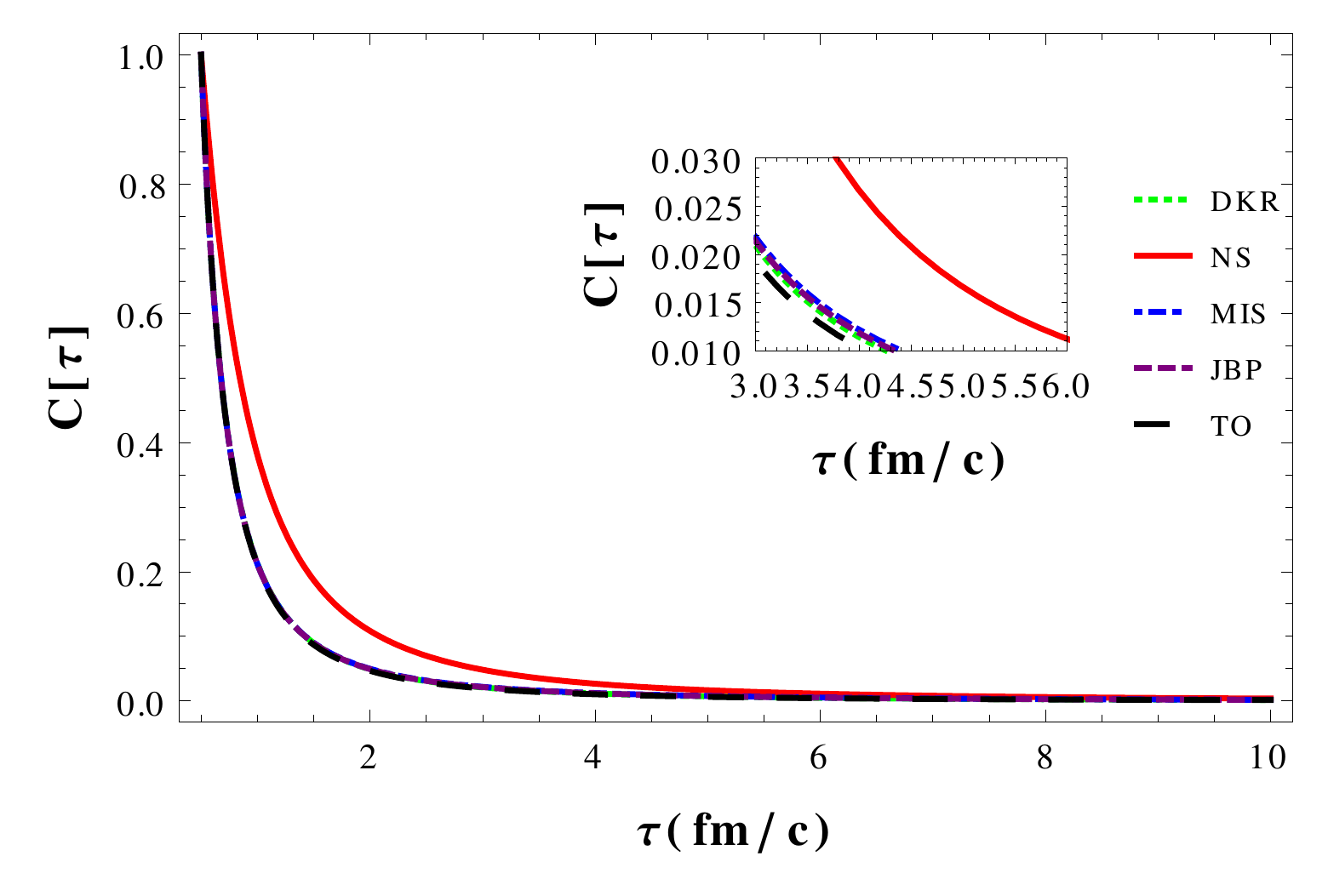}}\hspace{0.0cm}
\subfigure[]{\includegraphics[width= 8.0cm]{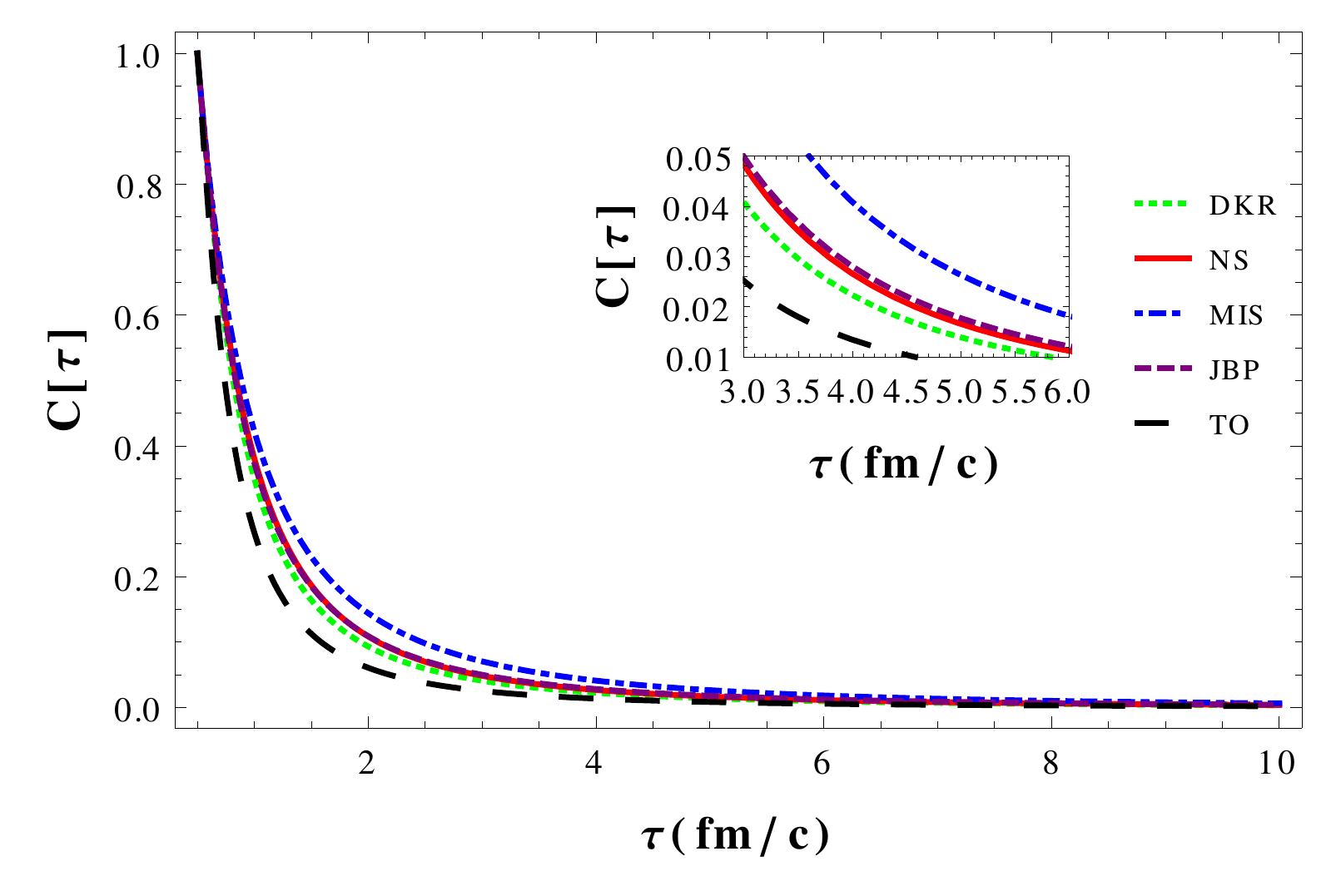}}\vspace{0.0cm}
\subfigure[]{\includegraphics[width= 8.0cm]{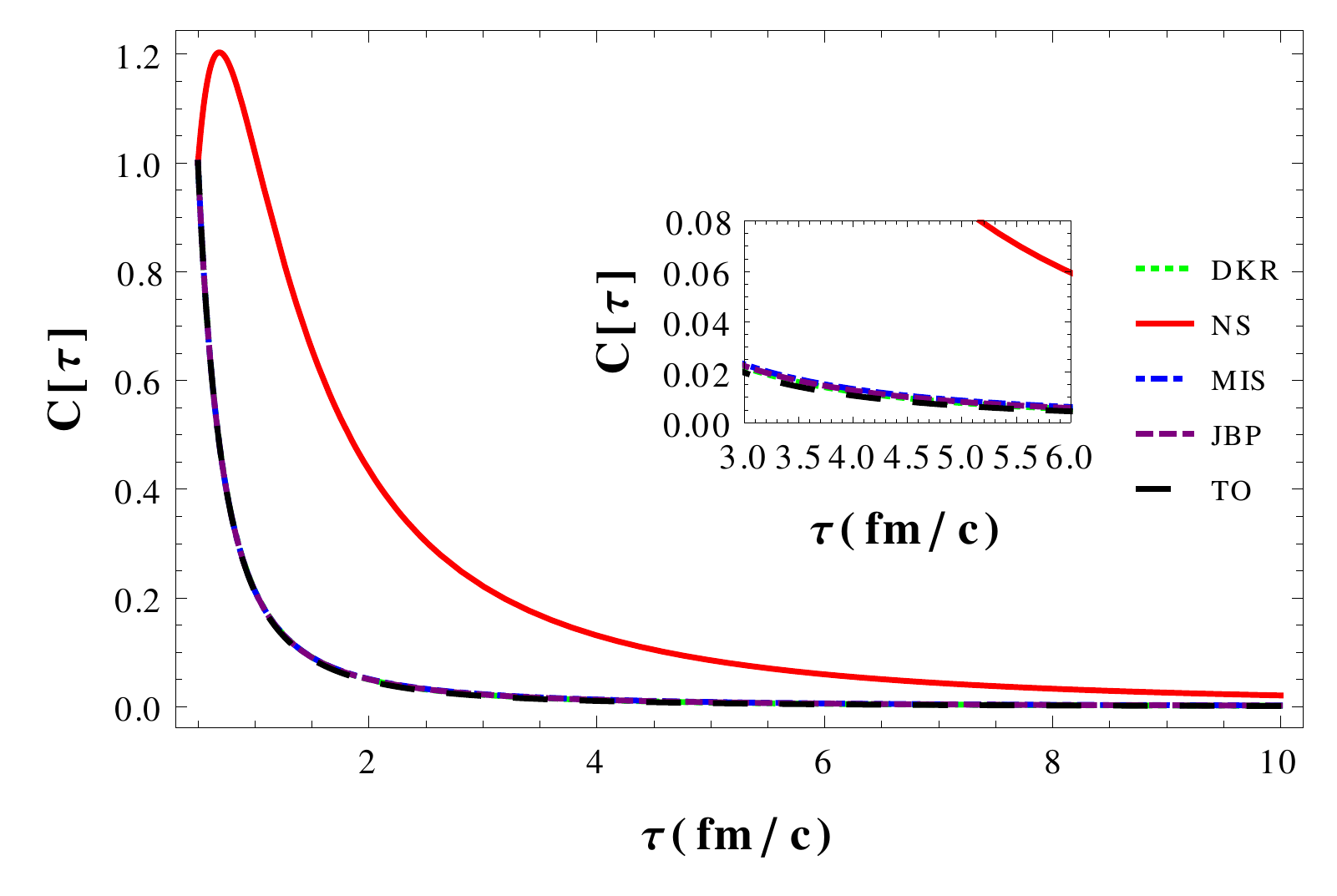}}\hspace{0.0cm}
\subfigure[]{\includegraphics[width= 8.0cm]{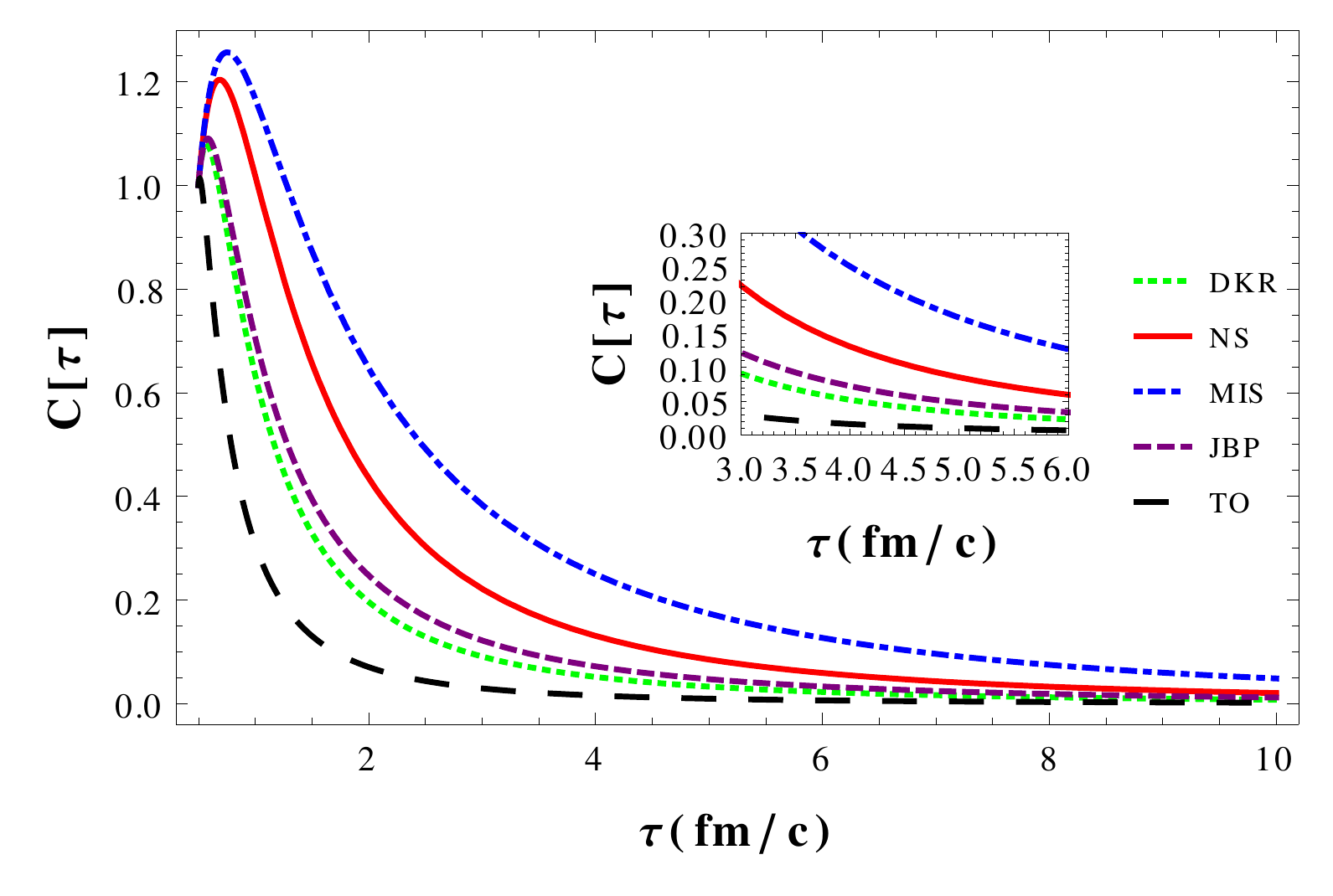}}\vspace{0.0cm}
\caption{ (2a), (2c), (2b), (2d), (2e) and (2f) show the time evolution of the function $C(\tau)_{[E]}$[see Eq.(\ref{96})]
with same initial temperature $T_i=310~MeV$. Where, [E] is corresponds to NS, MIS, DKR, JBP and TO hydrodynamics. Note 
that in all the figures the ratio of the 
viscosity to entropy density is kept same for all the Hydrodynamic approaches.
Fig.(2a), (2c) and (2e) corresponds to $\frac{\eta}{s}=0.08,\,\ 0.56,\,\ 1.60$ 
respectively with initial time $\tau_0=0.5$ fm/c and $\pi_0=0.0$ for all causal approaches. 
While Fig.(2b), (2d) and (2f) corresponds to the same $\frac{\eta}{s}$ and $\tau_0$ as in the former cases but 
with $\pi_0$ equal to Navier-Stokes initial value for all the hydrodynamic approches.}
\label{fig[2]}
\end{figure} 

There are two possible comparisons between the correlation
functions $C(\tau)_{[E]}$. 
In one such comparison the energy-independent cross-section $\sigma$ [see Eqns. (\ref{101}, \ref{102})] 
is kept same for all the different versions of the hydrodynamics\cite{rischke}.
Following Ref. \cite{huovinen,rischke}, one can write the viscosity
coefficient for the different models of hydrodynamics as $\eta_{DKR}=\frac{4T}{3\sigma}$,   
$\eta_{MIS}=\frac{6T}{5\sigma}=\eta_{JBP}=\eta_{TO}$ and $\eta_{NS}=0.8436\frac{3T}{2\sigma}$. 
Thus the relation between different $\eta$ are  given by the  scaling:
 $\eta_{MIS}=\eta_{JBP}=\eta_{TO}=9/10\eta_{DKR}$ and $\eta_{NS}=\frac{7.59}{8}\eta_{DKR}$.  
In the another way of comparing $C(\tau)_{[E]}$, the ratio $\eta/s$ is kept same 
for the different models of the hydrodynamics, while the  
$\sigma$ is varied for the different models.

 Figure (1) shows the case when the transport cross-section is kept same for all the
models of hydrodynamics.
The inset figure in all the diagrams 
shows the plots of correlation functions with better resolution in $\tau$ range 
between $3 fm/c$ to $6 fm/c$.   
 Cases (a-b), (c-d) and (e-f) corresponds to $\frac{\eta_{DKR}}{s}=0.08 ,0.56$ and $1.60$.
Values of $\eta/s$ for other models can be found using the scaling relation discussed above. 
The initial temperature $T_i$ and initial time $\tau_i$ are respectively chosen to
be $310~MeV$ and $0.5$fm/c.

One can notice for figures 1(a-b) that when $\eta_{DKR}/s$ is close to the minimum possible value ($1/4\pi$), 
all the correlation overlaps with each other. This is expected as all the viscous
hydrodynamics models should approach the ideal hydrodynamics limit when $\eta/s\approx 1/4\pi$.
Figures 1(c-d) corresponds to the case when $\frac{\eta_{DKR}}{s}=0.56$, i.e. almost seven times
larger than the most minimum value, the correlations only marginally differ from each other.
Overall difference between the correlation functions obtained using initial condition
$\pi=0$ and $\pi\neq 0$ is not significant. However, when $\pi=0$ case Navier-stoke correlation
slightly dominates over the correlation functions obtained using the causal models. While
for the case when the initial value of $\pi$ is same as Navier-Stoke value, it is the MIS correlation
function dominates over the other correlation functions.
Figures 1(e-f), correspond
to the case when $\eta/s$ almost twenty times larger than the minimum value. In figure 1(e) the
Navier-Stokes correlation first increases with time and then decreases. However, all the causal
models correlation decreases with time. Rise in the Navier-Stokes correlation can be attributed
to the unphysical behavior noted in Ref. \cite{BRW}. In this case it may be possible
to distinguish between the correlation function from the Navier-Stoke theory from the 
causal hydrodynamics models. However, the correlation function of the causal models overlaps
with each other. But when the Navier-Stokes value for the initial stress $\Pi_0$ is chosen
for the causal models, all the correlation function first increases with time and later
the plummet with time. This case can be considered to be unphysical as for all the
hydrodynamics models initially $\epsilon+p<\Pi$. The condition $\epsilon+p<\Pi$
violates the validity of the second order hydrodynamics.

Figure (2) corresponds to the case when the ratio of the viscosity coefficient to the entropy density
is kept same for all the five models of hydrodynamics. 
Cases (2a-2b), (2c-2d) and (2e-2f) respectively corresponds to the situation when
$\frac{\eta}{s}$ equal to $0.08 ,0.56$ and $1.60$. The initial temperature and the initial
times are kept same as in the case for figure (1). One can notice that as $C(\tau)$ in
Eq.(\ref{96}) remains same for all the hydrodynamical models, all the correlation
functions, starts at the same initial value. This was not the case in figure (1). 
Otherwise the general features about the correlation function remain same as in figure (1). 
Moreover, we have changed the values of initial temperature and initial time. 
In these cases also the general features of the correlation function remains similar to
those discussed in figure (1).

Finally we would like to discuss the importance our results. We first like to note 
that in the present work we have extended the formalism to calculate hydrodynamic 
fluctuations given in Ref\cite{landau1} to the relativistic causal theories. We have 
demonstrated that the form of the correlation functions in causal hydrodynamic theories 
remains same as in the relativistic Navier-Stoke case\cite{kapusta}. This result is not expected 
apriori, as the underlying hydrodynamic equations for the causal 
theories\cite{romatschke,mis,ajel,bhalerao,rischke} are very different than the 
Navier Stokes equation.
Equations ((\ref{27}-\ref{29}),(\ref{47}-\ref{49}), (\ref{57}), (\ref{65}-\ref{67}), (\ref{82}) and (\ref{85}))
can be employed to calculate the two particle correlators [see Ref\cite{kapusta}], which 
can be compared with the experimental 
data. However, this would require the solution of inhomogeneous (with noise term) hydrodynamical 
equations (of different types) in 3-dimension.
Further, in the present example we have dealt with boost invariant one 
dimensional flow. However, for a non-central heavy-ion collision, the vorticity can play 
a significant role\cite{wiedmann}. The presence of finite vorticity can cause the difference in the evolution 
in the correlation function for the different models of hydrodynamics remains to
be seen. One can notice from Eq.(\ref{69}) that vorticity can drive dynamics of the viscous stress
in DKR hydrodynamics. However this will require to solve hydrodynamical equation in $2+1$ or 
$3+1$ dimensions. This is at present, beyond the scope of this work.
Finally the numerical example that we have considered here, we plot the correlation function 
vs time. However, this numerical result can not be compared with the experimental data. But, 
this can give us some idea about how the correlation-function of different hydrodynamics compare with 
each other.
We find that the correlation functions obtained using various 
causal theories do not significantly differ from each other for a variety of values of initial 
conditions and $\eta/s$. However, the correlation function obtained using NS-theory can have unphysical 
behavior for higher values of $\eta/s$ and the NS-correlation function differ from the 
correlation functions obtained using the causal hydrodynamics.

\section*{5.  Conclusions}

We have studied fluctuations in various models of relativistic causal hydrodynamics.
We have found that the general properties of the dissipative part of the energy-momentum
tensor due to the viscosity and heat-flux play an important role in determining the Onsager
coefficients and the correlation functions. 
We find that the analytic form of the correlation 
functions remains same for all the causal hydrodynamics that considered 
here and do not depend explicitly on the relaxation time. Further our numerical investigations also suggest that the qualitative
behavior of the correlation functions for the various models of the causal hydrodynamics
remains similar to those of the Navier-Stokes theory  at least for a one dimensional boost-invariant
flow.

 {\it Note added:} 
 After this manuscript was prepared, we have found  in Ref\cite{murase} on arXiv that the authors have 
applied the fluctuation-dissipation relation to the relativistic viscous hydrodynamics with 
the memory effects. We have also found that in Ref\cite{clint} the authors have calculated 
hydrodynamic fluctuation for MIS Hydrodynamics. Ours and their results match with each other. 
In this work \cite{clint} the author has obtained dynamics of the noise-function, while
in our approach the noise-function is assumed to be given. However, we believe that one can obtain the noise
function dynamics from the arguments similar to the one given in Ref\cite{muronga} to obtain the 
dynamics of dissipative fluxes.

\section*{6.  Acknowledgements} The authors (AK, JRB) would like to thank to Prof. P. K. Kaw for 
helpful discussions. We would also like to thank the referee for his constructive criticism.


\begin{thebibliography}{99}

\bibitem{landau} L. D. Landau and E. M. Lifshitz, Statistical Physics: Part 1
(Pergamon, Oxford, 1980).

\bibitem{landau1} L. D. Landau and E. M. Lifshitz, {\it Statistical Physics: Part 2}
(Pergamon, Oxford, 1980).

\bibitem{salie} N. Salie, R. Wuffert, and W. Zimdahl, J. Phys. A: Math. Gen. {\bf 16},
  3533 (1983).
  
\bibitem{calzetta} E. Calzetta, Class. Quantum Grav. {\bf 15}, 653 (1998). 

\bibitem{geroch} R. Geroch, and L. Lindblom, Phys. Rev. {\bf D41}, 1855 (1990).

\bibitem{kapusta} J. I. Kapusta, B. Muller, and M. Stephanov, Phys. Rev. {\bf C85}, 054906 (2012).

\bibitem{kapusta1} J. I. Kapusta, J. M. Torres-Rincon, Phys. Rev. {\bf C86}, 054911 (2012).

\bibitem{hiscock} W. A. Hiscock, and L. Lindblom, 
Phys. Rev. {\bf D31}, 725 (1985); Ann. Phys. {\bf 151}, 466 (1983).

\bibitem{romatschke} P. Romatschke, Int. J. Mod. Phys. {\bf E19}, 1 (2010) (arXiv:0902.3663[hep-ph]).

\bibitem{mis} I. M\"{u}ller, Z. Phys. {\bf 198}, 329 (1967); W. Israel, Ann. Phys. {\bf 100}, 310 (1976); 
W. Israel, and J. M. Stewart, Ann. Phys. {\bf 118}, 341 (1979).

\bibitem{djou} D. Jou, J. Casas-Vazquez, and G. Lebon, Rep. Prog. Phys. {\bf 51}, 1105 (1988); {\it ibid}, {\bf 62}, 1035 (1999).

\bibitem{herrera} L. Herrera and D. Pavon, Phys. Rev. {\bf D64}, 088503 (2001); L. Herrera and D. Pavon, Physica {\bf A307}, 121 (2002).

\bibitem{muronga} A. Muronga, Phys. Rev. {\bf C69}, 034903 (2004). 

\bibitem{muronga1} A. Muronga and D. H. Rischke, arXiv:nucl-th/0407114.

\bibitem{heinz} U. Heinz, H. Song, and A. K. Chaudhuri, Phys. Rev. {\bf C73}, 034904 (2006); U. Heinz, arXiv:nucl-th/0512049.

\bibitem{berera} M. Bastero-Gil, A. Berera, R. Cerezo, R. O. Ramos, and G. S. Vicente, Jour. Cos. Astro. Phys. {\bf 1211}, 042 (2012). 

\bibitem{ajel} A. El, Z. Xu, and C. Greiner, Phys. Rev. {\bf C81}, 041901(R) (2010). 

\bibitem{bhalerao} A. Jaiswal, R. S. Bhalerao, and S. Pal, Phys. Rev. {\bf C87}, 021901(R) (2013).
 
 \bibitem{rischke} G. S. Denicol, T. Koide, and D. H. Rischke, Phys. Rev. Lett. {\bf 105}, 162501 (2010).

\bibitem{roma1} R. Baier, P. Romatschke, D. T. Son et al., J. High Energy Phys. {\bf 0804} (2008) 100.

\bibitem{muronga2} A. Muronga, Phys. Rev. {\bf C76}, 014909 (2007). 

\bibitem{bjorken} J. D. Bjorken, Phys. Rev. {\bf D27}, 140 (1983).

\bibitem{kouno} H. Kouno, M. Maruyama, and F. Takagi, Phys. Rev. {\bf D41}, 2903 (1990). 

\bibitem{huovinen} P. Huovinen and D. Molnar, Phys. Rev. {\bf C79}, 014906 (2009);  
Nucl. Phys. {\bf A830}, 475C (2009).

\bibitem{jaiswal} A. Jaiswal, arXiv:1302.6311[nucl-th].

\bibitem{BRW} R. Baier, P. Romatschke and U. A. Weidemann, Phys. Rev. {\bf C73}, 064903 (2003).

\bibitem{wiedmann} S. Florchinger, and U. A. Wiedemann, J. High Energy Phys. {\bf 11} (2011) 100.


\bibitem{murase} K. Murase and T. Hirano, arXiv:1304.3243[nucl-th].

\bibitem{clint} Clint Young, arXiv:1306.0472[nucl-th].



\end{thebibliography}
\end{document}